%% file: ms.tex
\documentclass[iop]{emulateapj}

\usepackage{color}
\usepackage{hyperref}
\hypersetup{colorlinks,citecolor=blue}
\newcommand{\HII}{H\,{\sc ii}}
\newcommand{\HI}{H\,{\sc i}}
\newcommand{\BrG}{Br\,{$\gamma$}}
\newcommand{\PaA}{Pa\,{$\alpha$}}
\newcommand{\sersic}{S\'ersic}
\newcommand{\Htwo}{$\rm H_2$}
\newcommand{\Vlsr}{$V_{\rm {\small LSR}}$}
\newcommand{\Nlyc}{$N_{\rm {\small Lyc}}$}

\slugcomment{{To Be Published in The Astrophysical Journal}}

\shorttitle {Nucleus of NGC\,6946 in Near-IR} \shortauthors{Tsai et al.}

\begin{document}

\title{The Circumnuclear Star Formation Environment of NGC\,6946: \\ \BrG\ and \Htwo\ Results from Keck Integral Field Spectroscopy}

\author{Chao-Wei Tsai\altaffilmark{1,2}, Jean L. Turner\altaffilmark{3}, Sara C. Beck\altaffilmark{4}, David S. Meier\altaffilmark{5,6}, and Shelley A. Wright\altaffilmark{7}} 

\altaffiltext{1}{Infrared Processing and Analysis Center, California Institute of Technology, Pasadena, CA 91125} 
\altaffiltext{2}{current address: Jet Propulsion Laboratory, California Institute of Technology, Pasadena, CA 91109; [email: \email{Chao-Wei.Tsai@jpl.nasa.gov}]}
\altaffiltext{3}{Department of Physics and Astronomy, UCLA, Los Angeles, CA 90095-1547}
\altaffiltext{4}{Department of Physics and Astronomy, Tel Aviv University, Ramat Aviv, Israel}
\altaffiltext{5}{Department of Physics, New Mexico Institute of Mining Technology, Socorro, NM 87801}
\altaffiltext{6}{National Radio Astronomy Observatory, P.O. Box O, Socorro, NM 87801}
\altaffiltext{7}{Dunlap Institute for Astronomy and Astrophysics, University of Toronto, Toronto, Ontario, Canada}

\begin{abstract}
We present a 3-dimensional data cube of the K band continuum and the \BrG, \Htwo\ S(0) and S(1) lines within the central 18\farcs5 $\times$ 13\farcs8 (520 pc $\times$ 390 pc) region of NGC\,6946. Data were obtained using OSIRIS, a near-infrared Integral Field Unit at Keck Observatory, with Laser Guide Star Adaptive Optics. The 0\farcs3 resolution allows us to investigate the stellar bulge and the forming star clusters in the nuclear region on 10 parsec scales. We detect giant \HII\ regions associated with massive young star clusters in the nuclear spiral/ring (R$\sim$30 pc) and in the principal shocks along the nuclear bar. Comparisons of the \BrG\ fluxes with \PaA\ line emission and radio continuum indicate $A_K\sim 3$, $A_V \sim 25$ for the nuclear star forming regions. The most luminous \HII\ regions are restricted to within 70~pc of the center, despite the presence of high gas columns at larger radii (R$\sim$200 pc). \Htwo\ emission is restricted to clouds within R$\sim$60 pc of the center, and in this sense resembles the distribution of the HCN line emission. We propose that gas-assisted migration of the young star clusters is contributing to the buildup of the nuclear bar and nuclear star cluster (R$< $30 pc) in this galaxy.

\end{abstract}


\section{Introduction}

The formation of high mass circumnuclear clusters has been observed in many nearby galaxies \citep[see review of][]{1997RMxAC...6....5H,1997RMxAC...6..165E}. In galactic centers, or in strong bars where dynamical timescales are short, the relatively older clusters that are visible have had time to move far from their birth places \citep{2009ApJ...703.1297E}. However, the youngest \HII\ regions are often deeply embedded in dusty clouds and do not have optical counterparts \citep[e.g. M82,][]{2009AJ....137.4655T}. To study star formation in situ requires knowing where the youngest, presumably embedded, clusters are, which requires radio or infrared identifications.

The fuel of star formation, molecular gas, is abundant in the central portions of spiral galaxy disks \citep{1978ApJ...223..803M,1991ARA&A..29..581Y}. Through tidal torquing and dissipation in spiral arms, gas is transported inward from the disk to the central few hundred parsecs at rates of a few tenths of a solar mass per year \citep{1992ApJ...384...72T,1997ApJ...482L.143R,2006ApJ...649..181S}. This inflowing gas has to either be accreted by a supermassive black hole, be expelled, or form stars. Despite enhanced gas columns in galactic centers, the nuclear environment does not guarantee high star formation rates: rapid rotation and strong nuclear tidal forces along the arms emerging from the central bar can intensify the shear across the molecular clouds. This is the likely cause of reduced star formation efficiency. However, there are abundant examples of massive young stars in galactic centers \citep{1958PASP...70..364M,1967PASP...79..152S,1996AJ....111.2248M,2000MNRAS.317..234P}, including our own Milky Way Galaxy \citep[see review by][]{1996ARA&A..34..645M}, showing that star formation can indeed occur in the galactic center environment.  Where and how does star formation take place in galactic centers?

NGC\,6946, a nearby barred spiral (SAB) galaxy, has currently active nuclear star formation, with $L_{\rm IR} \sim 2.2 \times 10^{9} ~L_{\sun}$ \citep{1996ApJ...467..227E}. At 5.9 Mpc \citep[1\arcsec = 28\,pc;][]{2000A&A...362..544K}, NGC~6946 harbors $10^{8} ~M_{\sun}$ of molecular gas within the central 170\,pc \citep[R = 3\arcsec,][]{2004AJ....127.2069M}. It is a compact, intense starburst with a star formation rate of $\sim 0.29 ~M_{\sun}~\textrm{yr$^{-1}$}$ within 8\arcsec\ from the center. The summary of physical parameters of NGC\,6946 are included in Table \ref{table:N6946}. The morphology of the gas in the inner $\sim$20\arcsec\ region ($\sim$ 450 pc diameter) indicates the presence of a strong nuclear bar \citep{2006ApJ...649..181S} that is of similar size and orientation to the nuclear bar evident in near-IR \citep{1998AJ....116.1221E}. NGC\,6946 does not have a distinct nuclear cluster in optical wavebands \citep{2002AJ....123.1389B}, unlike more than half of spiral galaxies \citep{1998AJ....116...68C,2002AJ....123.1389B}. Nor does it have dynamical signatures of a supermassive black hole \citep{2007AAS...211.6906K,2010ApJ...723...54K}. \cite{2007A&A...462L..27S} suggest that the bar is driving molecular gas inward so as to fuel star formation within 30 pc of the nucleus.

\input{table1.tex}

In this paper, we present the results of 0\farcs3 resolution Kn3-band (2.121 \micron--2.229 \micron) spectroscopic imaging of NGC\,6946 using the near-infrared integral field spectrograph OSIRIS (OH-Suppressing Infra-Red Imaging Spectrograph) with Keck Laser Guide Star Adaptive Optics (LGS-AO). The spatial resolution of 0\farcs3 corresponds to 8\,pc on the source. We consider the major features in continuum, \Htwo\ lines, and \BrG\ line in the central 18\farcs5 $\times$ 13\farcs8 region of NGC\,6946. We combine our data with the HST NICMOS images and sub-arcsecond CO maps from the literature to investigate the nuclear stellar bulge, stellar and recombination line extinction, molecular hydrogen distribution and excitation, and the formation of star clusters in the nuclear gas spiral structure. 

\section{Observations and Data Reduction}\label{section:obs}

We observed NGC\,6946 with OSIRIS \citep{2006SPIE.6269E..42L} on the Keck II Telescope on July 25 2008 (UT) under normal seeing conditions. OSIRIS is an integral field spectrograph working in the near infrared, designed to use adaptive optics for spectral imaging with the highest spatial resolution. The uncorrected optical seeing during the observation from the Subaru DIMM seeing monitoring service was 0\farcs8, which suggests seeing $\sim$ 0\farcs6 in K band. The spectral data cube was made in the Kn3 (K narrowband-3) filter with a built-in pupil for our 100 mas scale. The 0.1$\arcsec$/pix image scale results in a 0.25 nm/pix spectral scale, and $4\farcs8 \times 6\farcs4$ field of view. Aberration and diffraction effects under the 100-mas pupil result in a final spectral resolution of 2.5--3 pixels or $R=\Delta \lambda / \lambda \lesssim$ 3,400. The LGS-AO system \citep{2006PASP..118..297W} was used during the observation to correct for atmospheric aberration. The R=12.2 mag tip-tilt star (Strehl ratio of 0.40) at a distance of $55\arcsec$ from the data cube center provides a Strehl ratio of $\sim$ 9.6\% at  NGC\,6946's nucleus. The point spread function (PSF) after using the AO system is estimated to be $60\pm1$ mas on the tip-tilt star (at scale of $0.035\arcsec/pix$). Aberration and diffraction effects under the 100 mas pupil blur our PSF to 2.5 to 3 pixels (FWHM $\sim$ 0\farcs3). The final mosaic data cube consists of 9 pointings covering a region of $18\farcs5 \times 13\farcs8$ on the sky at position angle of $30\degr$. Integration time on each pointing was 300 seconds. 

Data reduction was done in \textsc{idl} with the OSIRIS pipeline software developed at UCLA \citep{2006SPIE.6269E..42L}. A custom routine was used to remove spurious bad pixels, defined as more than twice as bright as the strongest $Br\gamma$ emission in the cube. Data analysis and figure preparation are done in \textsc{idl} and \textsc{ds9}.

The telluric correction is made by using A0V star HD 195500 and G2V star HD 193193 that we acquired in the beginning of the night. The flux is calibrated to the 2MASS Ks band flux using two nearby bright stars with spectral types of A0 (V650 Cas; 2MASS Ks = 6.2 at $23^\textrm{h}48^\textrm{m}38^\textrm{s}.99$; $+64\arcdeg52\arcmin35\farcs5$) and G2 (Ks = 4.9 at $21^\textrm{h}13^\textrm{m}42^\textrm{s}.47$; $+64\arcdeg24\arcmin14\farcs1$) which were observed in the same Kn3 filter. They were intended to be telluric standards. However, with Keck AO they turned out to be resolved binaries (separation $\sim$ 0\farcs8 for V650 Cas, and 0\farcs4 for the other), and not suitable for telluric correction. Instead, we used these two binary systems for flux calibration with the following modifications. We applied the telluric correction obtained earlier in the night to observations of our flux calibrators and the NGC\,6946 data cube. On the flux calibrators, we apply the 2MASS Ks-band filter profile, and the estimated fractional contribution of the Kn3 window in the 2MASS Ks band (38\%). We convert from OSIRIS counts to flux by comparing the observed and the 2MASS Ks-band magnitude. The absolute uncertainty in the fluxes we obtain from OSIRIS counts is estimated to be 10\%. 

The wavelength is calibrated to the vacuum wavelength in the OSIRIS pipeline. Velocities reported in this paper are relative to Local Standard of Rest (\Vlsr) and are found using \textsc{iraf} routine \textsc{rvcorrect}. The uncertainty of the absolute velocity is $\sim$ 3 km/s, mainly reflecting the uncertainty of wavelength calibration\footnote{David Le Mignant, ``Note on OSIRIS Wavelength Calibration'', 2007. \url{http://www2.keck.hawaii.edu/inst/osiris/OsirisNote_dlm.pdf}.}.

The Kn3 band observation covers the \BrG\ line ($\lambda_{\rm rest} = 2.16612 ~\micron$), \Htwo\ 1-0 S(1) ($\lambda_{\rm rest} = 2.12183~\micron$), and \Htwo\ 1-0 S(0) ($\lambda_{\rm rest} = 2.22329~\micron$). The stellar continuum image is the median value of the line-free channels.  Detections of all of these emission features are clear except for \Htwo\ 1-0 S(0) which is marginally detected (4 $\sigma$). The overall noise level of the data cube is $\sim 1\times 10^{-15}$ $\textrm{W}/\textrm{m}^{2}/\textrm{\micron}/\textrm{arcsec}^{2}$ at spatial resolution of 0\farcs3. Near atmospheric absorptions, the noise level can rise by a factor of 3--5. 

Continuum-subtracted spectra integrated over the line emission region (central 5\farcs9 $\times$ 2\farcs9, P.A. = 125$\degr$) of the \Htwo\ 1-0 S(1), \BrG, and \Htwo\ 1-0 S(0) lines are shown in Figure \ref{figure:spec}. The continuum contributions to the line channels are estimated using the continuum emission in the consecutive line-free channels. We have clear detections of the \BrG\ line across 11 velocity channels centered at \Vlsr\ $\sim$ 50 km/s. The \Htwo\ 1-0 S(1) line strength is similar to the \BrG\ line in the central 5\arcsec\ region, and the profile slightly broader than the recombination line. The \Htwo\ 1-0 S(0) line is weaker than 1-0 S(1). It appears in our data at \Vlsr\ = 0 km/s, and could suffer from incomplete telluric and stellar absorption removal on the red wing of the line. Both \Htwo\ lines overlap with possible red giant/supergiant absorption features discussed by \cite{1996ApJ...467..227E}, and the noise is accordingly high. We do not detect \Htwo\ 2-1 S(1) as expected.

\input{fig1.tex}

We obtained \textit{HST NICMOS} images in F187N, F190N, and F160W bands of NGC\,6946's nucleus (proposal ID: 9360; PI: R. Kennicutt) from the \textit{HST} archive. We use the standard CALNICA and CALNICB pipeline products from Space Telescope Science Institute. The F160W data is used to representing the stellar continuum near the H-band. 
The \PaA\ image is produced by subtracting the off-line continuum emission in filter F190N from the F187N data. The preparation follows the procedures discussed by \cite{1999ApJS..124...95B}. 

For absolute astrometry, we use VLA radio point sources from \citet[][0\farcs4 in resolution]{2006AJ....132.2383T}. These radio sources, named source A--I in \cite{2006AJ....132.2383T}, are a mixture of thermal (\HII\ regions) and nonthermal (supernova remnants) emission which are distinguishable based on their radio spectral indices. Because of the complex radio morphology in the central 2\arcsec\ area, only 4 unconfused radio sources detected with signal-to-noise ratio of 4--10 at 6~cm can be used. Two of the four show \BrG\ counterparts. To improve the source registration, we acquired 18 cm (1.658 GHz) continuum data from the Multi-Element Radio Linked Interferometer Network (MERLIN) archive. The MERLIN observations were done November 22--29 1999 on the center of NGC\,6946 (proposal code: 99B/40; PI: A. McDonald), and during February 07--22 2005 on the supernova SN 2004et at $20^\textrm{h}35^\textrm{m}25\textrm{s}.40$; $+60\arcdeg07\arcmin17\farcs6$ \citep{2005MmSAI..76..565A}. Both data sets were reduced by the MERLIN archive pipeline. The calibrated ($u,v$) data were combined by the \textsc{aips} task \textsc{dbcon}, and were cleaned and mapped using \textsc{imagr} in \textsc{aips}. The synthesized beam size of the naturally-weighted map is 0\farcs18 $\times$ 0\farcs15 (P.A. = 75$\degr$), comparable to the OSIRIS \BrG\ spatial resolution ($\sim$ 0.3\arcsec). The flux density rms in the map is 0.056 mJy/beam, and the SN ratio better than 8 in 5 compact sources. We revised the coordinates of radio continuum source C identified at 6~cm \citep{2006AJ....132.2383T} but confused by extended emission in the lower resolution VLA maps. The revised location of source C (renamed as C$^{\prime}$) is listed in Table \ref{table:flux_compact}.

The final absolute astrometry of the OSIRIS data, HST NICMOS \PaA\ and F160W images is registered to the astrometry of VLA \citep[][at 2~cm and 6~cm]{2006AJ....132.2383T} and MERLIN radio maps. We use the compact sources detected at both radio and near-IR wavelengths for registration, such as compact \HII\ regions in the NICMOS field and compact radio continuum (CRC) sources B, D, E, and H.   Cross-matching positions of VLA/MERLIN radio continuum and near-IR peaks gives absolute astrometry better than 0\farcs1.

\section{The Kn3 Band Stellar Continuum Emission}\label{section:Kn3_continuum}

NGC\,6946 is known to have an infrared stellar nuclear bar 8\arcsec\ long at position angle 140$\degr$ \citep{1998AJ....116.1221E}. The K-band stellar continuum in the OSIRIS cube is detected ($> 10 \,\sigma$) over a total of 433 channels. Figure \ref{figure:Kn3_obs}a shows the Kn3 band continuum image; the 7\arcsec\ by 5\arcsec\ emission region peaks at R.A. $= 20^{h}34^{m}52^\textrm{s}.29$, decl. $= +60\arcdeg09\arcmin14\farcs0$, on the southeastern side of the dynamical center (listed in Table \ref{table:N6946}). This near-IR continuum bar is 200 pc $\times$ 140 pc  (projected) in size. 

We also resolve a few compact near-IR sources in our Kn3 continuum image (see Figure \ref{figure:Kn3_obs}a), which are also detected in the HST NIC2 F160W image. One bright source is on the northeastern side of the main stellar continuum emission at R.A. $= 20^\textrm{h}34^\textrm{m}52\textrm{s}.407$, decl. $= +60\arcdeg09\arcmin17\farcs05$, outside the high extinction region ($A_V \lesssim$ 2~mag; see \S\ref{section:stellar_extinction}). The flux density of this source is $1.9 \times 10^{-16}~W/m^2/\micron$, or $m_{{\rm K}}^{\rm Vega} = 15.8$ ($m_{{\rm K}}^{\rm AB} = 17.7$), comparable to that of a globular cluster or a foreground dwarf satellite galaxy at the distance of NGC\,6946. However, other Kn-band sources we identified are close to the high extinction area discussed in \S\ref{section:stellar_extinction}. It is not clear if they are clusters or the results of localized patchy obscuration.

\input{fig2.tex}

\subsection{Extinction and Dereddening; Stellar Population}\label{section:stellar_extinction}

The extinction toward the stellar continuum is estimated by comparing the HST NIC2 F160W band image and our Kn3 stellar continuum map. We assume the F160W-Kn3 color of the stellar component to be constant at $-0.2$ magnitude, the median value for the H-K color of a stellar population with continuous 0.1 $M_{\sun}\;yr^{-1}$ star formation over $>$ 100 Myr \citep[from STARBURST99;][]{1999ApJS..123....3L}. Then we calculate the extinction from the excess of the H-K color with the assumptions of foreground screening geometry and an extinction law \citep{1989ApJ...345..245C,1997AJ....113..162C} of $A_{V} = 34.4 \times \log\left(\frac{F_{\lambda}^{Kn3}}{F_{\lambda}^{F160W}}\right) + 6.6$. 

The extinction map in $A_V$ is shown in Figure \ref{figure:Av_conti}. The extinction map is morphologically in excellent agreement with the PdBI high resolution CO(2-1) map \citep{2007A&A...462L..27S} as shown in the figure. We deredden our data using the $A_{V}$ map and the assumption that $A_{Kn3}:A_{V} = 1.157:10$, from the extinction law of \cite{1989ApJ...345..245C}. The dereddened Kn3 band stellar emission is shown in Figure \ref{figure:Kn3_obs}b. However, even in the near-infrared, extinction is still present. We cannot meaningfully measure extinctions $A_{Kn3}\gtrsim1$ within the source in this way. In the center of NGC~6946, \citet{2006ApJ...649..181S} estimate  that on average $\rm N_{H_2}\sim 10^{23}~\rm cm^{-2}$ and $A_{V} \sim 125$ (corrected to $X_{\rm CO}$ from \cite{2004AJ....127.2069M}), or $A_{ K} \sim 15$. Although the $A_{V}$--$I_{CO}$ conversion could involve some uncertainty, this substantial difference suggests either patchiness in the molecular cloud coverage or that a significant amount of molecular gas is on the far side of the nuclear bulge along the line-of-sight in central 2\arcsec\ region. This picture is consistent with a gas ring \citep{2007A&A...462L..27S}.

\input{fig3.tex}

The stellar emission clump (the peak in the Kn3 band image) on the southeastern side of the center and another emission complex (near $20^\textrm{h}34^\textrm{m}52^\textrm{s}.13$; $+60\arcdeg09\arcmin15\farcs0$) on the northwest in Figure \ref{figure:Kn3_obs}a are far less prominent in the dereddened map. The position angle of the dereddened continuum emission does not exactly align with the observed (pre-dereddened) image, which had P.A. = $135\degr$, nor with the small stellar bar reported by \citet[P.A. = $140\degr$]{1998AJ....116.1221E}. The difference of 10$\degr$ -- 20$\degr$ in revised position angle is small, but could affect the dynamical modeling.

The dereddened stellar continuum flux density inside the central 4\arcsec\ aperture (central $\sim 100$~pc is $3 \times 10^{-16}~\textrm{W/m$^2$/\micron}$, or $\sim 0.9 \times 10^{6}\,L_{\sun,K}$. The sub-arcsecond CO line studies \citep{1990ApJ...355..436I,2006ApJ...649..181S} indicate a dynamical mass of $3 \times 10^{8}~M_{\sun}$ in the same region. Gas here is abundant, but not dynamically significant; molecular gas constitutes $\sim$10\% of the total mass within the central 120 pc \citep{2004AJ....127.2069M,2006ApJ...649..181S}. This high $M/L_{K}$ ratio enclosed within 2\arcsec\ radius supports the existence of a young population with an age of 0.1 Gyr or less. In fact, \cite{1996ApJ...467..227E} detected absorption features against the stellar continuum which match those of late type supergiants in NGC\,6946's nucleus. These features are also seen marginally in the OSIRIS data. This suggests that the K band emission is likely dominated by stars significantly younger than 100 Myr in the very center. 

\subsection{S\'ersic Profile}\label{section:sersic_model}

The stellar light profile can be described by a \sersic\ function:
\begin{equation}
I(R) = I_{e}\exp\left\{-b_{n}\left[\left(\frac{R}{R_{e}}\right)^{1/n} - 1\right]\right\},
\end{equation}
where $I(R)$ is the intensity at radius $R$, $I_{e}$ is the intensity at the effective radius $R_{e}$ in which 50\%\ of the light is enclosed, and $b_{n}$ is a constant for a given S\'ersic index $n$. The optical profile of NGC 6946's center shows no sign of a classical bulge \citep[with $n$ close to 4, i.e., De Vaucouleurs' profile; ][]{2010ApJ...723...54K}. Despite the high extinction across the region at optical wavelengths, a light profile of $n \sim 1.0$ at NGC\,6946's nucleus is determined by the seeing-limited ground-based V, R, and I band images \citep{2001A&A...367..405P}, and of $n \sim 0.92$ from work by \cite{2010ApJ...723...54K} using ground-based imaging and HST imaging. The low $n$ suggests a nuclear pseudo-bulge similar to that of the Milky Way \citep[$n$ = 1.0--1.3 at 2.4\micron;][]{1991ApJ...378..131K,2001ApJ...563L..11G,2007ApJ...655...77G,2010ApJ...723...54K} and late type galaxies in general \citep[see Section 4.2 of ][]{2004ARA&A..42..603K}. Based on the relation of S\'ersic index and mass of black hole from \cite{2007ApJ...655...77G}, one would expect the massive black hole at center of NGC\,6946 to be $\sim 2 \times 10^{5}~M_{\sun}$, close to the mass limit placed by a dynamical study in the optical \citep[$\sim 10^{5}~M_{\sun}$;][]{2007AAS...211.6906K}. 

With the leverage of having high resolution beyond the limitation of seeing, and less-extinction-affected line-free near-IR continuum images from the OSIRIS data cube, we re-examine the light profile of NGC\,6946's nucleus in near-IR. We use \textsc{galfit} \citep{2002AJ....124..266P} to model the S\'ersic profile of the de-reddened Kn3 stellar continuum map of the central $\sim 7\arcsec$ region. We included only the area with $> 1 \sigma$ detection. The best fit model is shown in Figure \ref{figure:Kn3_obs}c. Our model yields a \sersic\ index of $\sim 0.9$, consistent with the results from previous work at optical wavelengths \citep{2001A&A...367..405P,2010ApJ...723...54K}. 

The center of the stellar distribution in NGC\,6946 is at R.A. $= 20^\textrm{h}34^\textrm{m}52\textrm{s}.253$, decl. $= +60\arcdeg09\arcmin14\farcs10$. The uncertainty of the center is estimated to be $\sim 0\farcs1$. The best-fit model has a S\'ersic index of 0.91, position angle 153$\degr$, and 0.69 major-to-minor-axis ratio (Table \ref{table:N6946}). The effective radius $R_{e} = 2\farcs18$ ($\sim$ 65 pc) of our \sersic\ model is significantly smaller than the \sersic\ radius of 3\farcs5--3\farcs7 reported by \cite{2001A&A...367..405P} using seeing-limited V-, R-, and I-band images. This is probably due to our higher resolution and lower extinction.

The \sersic\ center of the main stellar pseudo-bulge does not coincide with the X-ray, infrared (2MASS), or radio continuum peaks \citep{1983ApJ...268L..79T,2003ApJ...588..792H,2004AJ....127.2069M}, nor with the CO(2-1) line dynamical center \citep{2006ApJ...649..181S}. The S\'ersic center is offset from the nearby radio continuum sources and the \BrG\ peak by $\sim$ 0\farcs15, which is significant compared with the absolute astrometry uncertainty of 0\farcs08.

\input{fig4.tex}

The inclination of NGC\,6946 of $40\degr$ at P.A = 242$\degr$ \citep{2007AJ....134.1827C} implies that the major-to-minor-axis ratio of 0.69 cannot be a projection effect, ruling out the possibility that the stars in the center are distributed in a perfect disc or a sphere. The S\'ersic ellipse of this nuclear pseudo-bulge is almost perfectly perpendicular ($89\degr$) to the position angle of the galaxy determined from H\,\textsc{i} kinematics \citep{2007AJ....134.1827C}. This result is consistent with the 210-pc-long stellar ``minibar'' perpendicular to an 1.9 kpc large-scale stellar oval distortion \citep{1979ApJ...229..111L,1984ApJ...278..564D,1986ApJ...303...66Z,1995ApJ...452L..21R,1998AJ....116.1221E} in near-IR and mid-IR images. The minibar structure has an exponential radial profile beyond the high extinction region (3\arcsec), and the width/length ratio is 0.6, significantly larger than that of a typical bar \citep[ratio $\sim$ 0.2;][]{1998AJ....116.1221E}.

\section{\Htwo\ $v$ = 1-0 S(1) and 1-0 S(0) Line Emission}\label{section:H2}

Pure rotational \Htwo\ emission in the mid-infrared from NGC\,6946 has been detected in ISO SWS observations \citep{1996A&A...315L.145V}, and by Spitzer \citep{2007ApJ...669..959R}. In the near-infrared, the lower vibrational transitions ($v < 2$) of NGC\,6946's nucleus have been investigated by different groups \citep{1988MNRAS.234P..29P,1996ApJ...467..227E,2004ApJ...609..692P,2004ApJ...601..813D}. 

The infrared vibration-rotation transitions of $H_2$ can be excited by shock heating of the molecular gas, a process we will refer to as {\it thermal} or collisional excitation, or by {\it fluorescence} in a strong FUV (far ultraviolet) radiation field. In FUV-excited gas the higher level vibration transitions ($v > 2$) will be stronger than in collisionally excited gas. Lines from these  $v > 2$ levels are weak and we did not detect any in NGC\,6946. Here we image two lines of \Htwo\ with OSIRIS, the $v$ = 1-0 S(1) and S(0) lines.

\Htwo\ $v$=1-0 S(1) has a rest wavelength 2.12183 \micron, which is at the edge of our wavelength coverage. The \Htwo\ emission is strongly confined to a region close to the dynamical center (Figure \ref{figure:H2_map}). Previous Fabry-P{\'e}rot spectroscopic measurements with a 11\farcs1 aperture on the region at similar spectral resolution show that the 1-0 S(1) line profile covers \Vlsr\ = -60 to 250 km/s \citep{2004ApJ...609..692P}, thus the anticipated missing flux due to our wavelength coverage limit at $\lambda_{\rm observed} < 2.121$ \micron ($\sim V_{\rm rest} < -100$) is small. We estimate the missing amount of \Htwo\ line emission to be $<$10\%, assuming a similar line profile for the integrated \Htwo\ line in the inner 5\arcsec~ (Figure \ref{figure:spec}a). A total flux of $5.1\pm0.9 \times 10^{-17}~W/m^2$ is found from our observed field in the first 12 channels (\Vlsr\ = -95--271 km/s) across a $\sim$ 6\arcsec $\times$ 6\arcsec region. Our S(1) flux is consistent with the flux obtained by \citet{2004ApJ...601..813D} with an IFU, but more than twice the results of \citet{1988MNRAS.234P..29P} and \citet{2004ApJ...609..692P}.  

In Figure \ref{figure:H2_map}, the integrated \Htwo\ S(1) line map shows the emission confined to the inner 5\arcsec\ ($\sim$ 150 pc), distributed toward the center and westward. Although, like the CO(2--1), the H$ _2$ emission peaks within the central  2\arcsec\ (Figure \ref{figure:H2_map}b), beyond that 50 pc radius the CO and H$_2$ show little spatial correlation. There is no detectable \Htwo\ emission toward the outer CO peaks located at projected radii of $\gtrsim$4\arcsec\ (120 pc) from the center. By contrast with CO, there is excellent spatial agreement between \Htwo\ S(1) emission and emission of the dense gas tracer HCN \citep{2007A&A...462L..27S}. HCN, often observed to be enhanced in galactic centers \citep{1997ApJ...478..162H,1997ApJ...484..656P}, is also confined to the very center of the galaxy between the CO peaks.

The \Htwo\ spectra at the \sersic\ center in a 0\farcs3 aperture are shown in Figure \ref{figure:H2_spec_center}. Both lines have similar dispersion (observed $\sigma_V \sim 140\pm30~ \textrm{km/s}$) and are relatively bluer than the systemic velocity at \Vlsr\ = 50 km/s. The total fluxes of the 1-0 S(1) and 1-0 S(0) line are $6\pm0.4 \times 10^{-19}~\textrm{W/m$^{2}$}$ and $3\pm0.4 \times 10^{-19}~\textrm{W/m$^{2}$}$, respectively. For fluorescent \Htwo\ molecules the branching ratios of vibration-rotation states in cascade give a \Htwo\ 1-0 S(0) to 1-0 S(1) line intensity, $R_{01}$, close to 0.5. In the thermal or collisionally excited case, $R_{01} \sim 0.2$ \citep{1987ApJ...322..412B,1988MNRAS.234P..29P,1994ApJ...427..777M,2004ApJ...609..692P}. The line ratio $R_{01}$ of 1/2 at the S\'ersic center would imply that the emission there is FUV excited. However, the S(0) line is only marginally detected ($\sim 4\,\sigma$), and also suffers telluric OH absorption and giant/supergiant star absorption lines on the red wing of the line (Figure \ref{figure:spec}c), while some flux is missing from the observed S(1) line as noted above. So $R_{01}$ has large uncertainties. 
 
\input{fig5.tex}

Over the entire region, we detect a total flux of $1.0\pm0.3 \times 10^{-17} \textrm{W/m$^2$}$ in \Htwo\ 1-0 S(0), substantially smaller than $2.6\pm0.7 \times 10^{-17} \textrm{ W/m$^2$}$ reported by \cite{1988MNRAS.234P..29P} for data taken with a 19\farcs6 FWHM, low spectral resolution photometer-spectrometer. We measure an integrated flux of \Htwo\ 1-0 S(0) for the emission region that is about 1/5 of that in 1-0 S(1) line, suggesting that the \Htwo\ here is thermally (i.e. collisionally) excited. This result is consistent with high spectral resolution, 2\farcs34 long-slit measurements on the NGC\,6946's nucleus by \cite{1996ApJ...467..227E} in which the two lines are measured in a single spectrum. In summary, though uncertainties are large it appears that \Htwo\ is collisionally excited everywhere except possibly a small region toward the very center of the galaxy. We discuss the \Htwo\ excitation later in connection with the \BrG\ kinematics. 

\section{Brackett $\gamma$ Line Emission, The Nuclear Starburst, and Star Cluster Formation}\label{section:BrG}

The central few hundred parsecs of NGC\,6946 has long been recognized as a site of active star formation \citep{1979ApJ...229..111L,1980ApJ...235..392T,1983ApJ...268L..79T,1993AJ....106..948D,1996ApJ...467..227E}. The \BrG\ line from \HII\ regions excited by young stars is the strongest feature from the galaxy in the Kn3 band. In this section, we consider the features of \BrG\ emission morphologically and spectroscopically. 

The integrated \BrG\ line map is presented in Figure \ref{figure:BrG_PaA}. Two strong emission peaks separated by $\sim$ 1\farcs5 (42 pc) east-west flank the central emission complex. Fainter and more extended emission follows the general east-west trend. A linear structure at P.A. = 30$\degr$ around R.A. = $20^{h}34^{m}53^{s}$, decl. $= +60\arcdeg09\arcmin12\arcsec$--$16\arcsec$ (east of the main CO central bar--arm complex, and parallel to the eastern arm; see \cite{2007A&A...462L..27S}) is also marginally detected. 

\input{fig6.tex}

The \BrG\ emission is asymmetric, somewhat stronger on the western side (Figure \ref{figure:BrG_PaA}). Elongated filaments are present along the eastern dust lane that are absent in the west. Another elongated component (indicated by the arrows with dashed lines in Figure \ref{figure:BrG_PaA}) of 6\farcs5 in length at P.A. = 165$\degr$ across the S\'ersic center is seen in the HST \PaA\ map. The sensitivity of our measurement did not permit us to recover the southern part of the \PaA\ structure, but the northern part is visible.

Channel maps of \BrG\ line emission are displayed in Figure \ref{figure:Chan_map}. We applied two-dimensional spatial smoothing with a 3-point Hanning kernel. Each channel covers 0.25 nm in wavelength, or $\sim$ 35 km/s in velocity. The 100 mas pupil blurs our velocity resolution to $\sim$90 km/s, so the channels are not independent. The line is centered at \Vlsr\ = 58~km/s, consistent with the systemic velocity of \Vlsr\ 50--60 km/s reported by \HI\ and CO studies \citep{1988PASJ...40..511S,1990ApJ...355..436I,1990A&A...234...43C,2004AJ....127.2069M,2006ApJ...649..181S,2007AJ....134.1827C}. The line emission is blueshifted toward the east due to galactic rotation, with a total width of 7 channels ($\Delta v \sim$ 250 km/s). The central emission complex has a velocity range of \Vlsr\ = -115~km/s to 231~km/s.

\input{fig7.tex}

We recover a total \BrG\ flux of $2.8\pm0.3 \times 10^{-17} W/m^2$, which is significantly (35\%--45\%) smaller than fluxes reported for low spectral resolution in a 7\farcs2 \citep{1990ApJ...349...57H} and a 19\farcs6 aperture \citep{1988MNRAS.234P..29P}. Seeing-limited integral field spectrograph (IFS) study of a $5\farcs4 \times 9\farcs4$ region \citep{2004ApJ...601..813D} finds 2.5 times more flux than we do, although their absolute flux calibration is uncertain. Our result agrees with the narrow band measurements by \cite{1996ApJ...467..227E}. We discuss in \S\ref{section:Av_line} that our fluxes agree with predictions based on the radio free-free fluxes and extinction based on comparison with \PaA, so while we do not understand the discrepancies with these other measurements, there seems to be a consistency with the radio fluxes.

We detect strong \BrG\ emission objects coincident with CRCs B, D, E, and H discovered in \citet[][]{2006AJ....132.2383T}. The measured flux density of \BrG\ over the regions of radio continuum objects is listed in Table \ref{table:flux_compact}. We discuss these compact objects in \S\ref{section:compact_objects}.

\input{table2.tex}

\subsection{\BrG\ Line Profiles and the Distribution and Kinematics of Ionized Gas}\label{section:BrG_line_fitting}

We fit the \BrG\ spectra with single gaussians, with results shown in Figure \ref{figure:BrG_fitting}a--c. The fitting procedure is performed on each 0\farcs1 pixel of the \BrG\ data cube which is convolved to a 0\farcs3 PSF. The central velocity and dispersion of regions with high S/N can be determined better than our nominal $\sim$90~km/s resolution. The residual of the fits is $\sim0.9 \times 10^{-16}~W/m^2/\micron/arcsec^2$, or $\sim 1 \sigma$ of the observed data cube, spatially uncorrelated. 

\input{fig8.tex}

The spatial  variation of the \BrG\ line peak amplitude is shown in Figure \ref{figure:BrG_fitting}a. This presentation suppresses faint, broad \BrG\ features so narrow line and bright compact objects and filaments can stand out.

In Figure \ref{figure:BrG_fitting}b, we map the central velocity of the (\Vlsr) of \BrG\ line. In regions of low emission the velocity uncertainty is $\sim$ 100~km/s and the values are scattered; in strong emission regions, where the uncertainty is 10--30 km/s, a clear rotation pattern emerges. Velocities shift from \Vlsr\ $\sim$ 30~km/s on the east to \Vlsr\ $\sim$ 90 km s$^{-1}$ in the west. Emission from the center is bluer than both the gas immediately surrounding it and the systemic velocity by 20--30 km s$^{-1}$.

The \BrG\ line has a smaller gradient than that of the CO which shows a well-regulated nuclear spiral pattern driven by the nuclear bar \citep{2006ApJ...649..181S}. The velocity gradient of the ionized gas is 17~km/s/arcsec over 2\arcsec\ at position angle of 150\degr, corrected for the 40$\degr$ inclination of the galaxy, while the CO gradient over the same region is more than 50 km s$^{-1}$ arcsec$^{-1}$ \citep[Figure 10 of ][]{2006ApJ...649..181S}. 
The enclosed virial mass inside the \BrG\ emission nodes at 2\arcsec, found from these velocities, is about $2 \times 10^7~M_{\sun}$, while the stellar mass is $2 \times 10^{8}~M_{\sun}$ or more (see discussion in \S\ref{section:stellar_extinction}). The discrepancy of the velocity gradient between \BrG\ and CO and the underestimated $M_{\rm vir}$ suggest that the lines originate from different regions along the line-of-sight, higher velocity components in \BrG\ are missing due to extinction, and/or they are not co-moving. The motion could be non-axisymmetric. If the motion of ionized gas is predominantly lateral, that is, in the plane of the sky rather than in the line of sight, then the projected velocities we observe will be low compared to the true (space) velocities. The non-co-moving scenario could imply that the \HII\ regions are dynamically detached from the parent CO gas clouds, and eventually will physically separate themselves from the natal clouds with time (further discussion in \S\ref{section:SCF_nuclear_ring}). 

Figure \ref{figure:BrG_fitting}c shows the best-fit gaussian dispersion of the line. The fitting uncertainty in dispersion is less than 10 km/s. The dispersion is $\sim$ 50 km/s -- close to the resolution of OSIRIS -- across the emission region, significantly increasing to $\sim$ 80 km/s toward the central 1\arcsec, and plateauing around the S\'ersic center. In addition, the central velocity of the \BrG\ line decreases by 10 km/s--20 km/s from the 1\arcsec\ radius toward the center. Both of these phenomena indicate strong star formation within a nuclear ring \citep{2002MNRAS.329..502M}, where we use the term ``nuclear ring'' to refer to the circularly moving molecular gas discussed by \cite{2007A&A...462L..27S}.

\subsection{Extinction of Near-IR Hydrogen Recombination Line(s)}\label{section:Av_line}

We can estimate extinction to the star-forming regions from our \BrG\ flux by comparing to other recombination lines and to cm-wave and mm-wave free-free fluxes. A \PaA\ image was obtained from the HST archive, with proper continuum subtraction and flux conversion. The intrinsic \PaA/\BrG\ line ratio is 12.2 for electron density $n_e = 10^{4}~cm^{-3}$ and electron temperature $T_e = 5,000~K$ gas \citep{1995MNRAS.272...41S}. This ratio does not vary significantly in the range of $n_e$ and $T_e$ for \HII\ regions. The observed \BrG/\PaA\ line ratio will be higher than this in the presence of extinction when the simple screen extinction geometry is assumed for these compact sources. We find the reddening between \BrG\ and \PaA, adopting the extinction law of \cite{1994ApJ...429..582C}, of $A_V = 32.7 \times (A_{Pa\alpha} - A_{Br\gamma})$. The extinction map of the ionized gas obtained from \PaA\ is shown in Figure \ref{figure:BrG_fitting}d. The extinction found from the near-IR hydrogen recombination lines is $A_V \sim$ 15--30 ($A_{Br\gamma} \approx A_K \sim $ 1.7--3.5) over the \BrG\ emission region. The $A_K > 1$ indicates that the \PaA/\BrG\ line ratio is saturated. There is no sign of higher extinction toward the center of the galaxy and no correlation with A$_{V}$ determined from the continuum. This is an indication that much of the extinction is internal to the \HII\ regions so estimating extinction for the recombination lines from the stellar continuum is inappropriate, as noted by \cite{1994ApJ...429..582C}.

Free-free fluxes in the microwave region are unaffected by extinction and are proportional to the dereddened H recombination line fluxes. From the observed 2.7mm continuum flux of $8.3\pm 1$~mJy \citep{2004AJ....127.2069M}, which gives $N_{Lyc} \sim 5.3 \times 10^{52}$ sec$^{-1}$, and using
\begin{equation}\label{eq:Nlyc_from_BrG}
N_{\rm Lyc} = 0.38 \times 10^{49}~sec^{-1} \left( \frac{D}{Mpc} \right)^2 \left( \frac{S_{Br\gamma}}{10^{-18}~W/m^2} \right),
\end{equation}
for $n_e = 10^{4}~cm^{-3}$ and $T_e = 5,000~K$, we predict a derreddened \BrG\ flux of $4.6 \times 10^{-16}~W/m^2$, and thus an extinction, $A_K\sim 3.0$. This is close to the \PaA-derived dereddened value ($3.3 \times 10^{-16}~W/m^2$ and $A_K\sim 2.7$). This is also within 20\% of the $N_{Lyc} \sim 4.6 \times 10^{52}$ sec$^{-1}$ value of \cite{2010ApJ...709L.108M} based on 1.3\,cm continuum fluxes for an estimated 50\% synchrotron contamination. 

These independent fluxes are all in rough agreement that within the central 200 pc of NGC\,6946, $N_{Lyc} \sim 5\pm 1 \times 10^{52}$ sec$^{-1}$ and $A_K\sim 3$, $A_V \sim 25$. The \BrG\ lines detect only $\sim$5-10\% of the ionized gas. However, the extinction appears to be internal to the \HII\ regions, in which case the \BrG\ emission is still a good tracer of overall \HII\ region  morphology and kinematics, although not the \HII\ region cores.

\subsection{Compact Line Emission Sources: Embedded Young Star Clusters}\label{section:compact_objects} 

Nine compact radio continuum (CRC) sources ($\lesssim$ 15~pc) are identified in sub-arcsecond resolution VLA radio maps of NGC\,6946 \citep{2006AJ....132.2383T}. These CRC sources are overlaid on the \BrG\ amplitude map in Figure \ref{figure:RC_BrG_spec}a. Radio \HII\ regions B, D, E, and H coincide with four \BrG\ peaks, and are used for the image registration. Among the remaining five CRC objects, the declining radio spectrum source A has no associated \BrG\ peak. It is possibly a supernova remnant (SNR). The spectral indices and natures of G and F are unknown, and their lack of \BrG\ make it probable that they are non-thermal SNRs. The other six CRC sources are giant \HII\ regions.

\input{fig9.tex}

The \BrG\ spectra of all the CRC sources (in 0\farcs3 regions, circled in Figure \ref{figure:RC_BrG_spec}a) are shown in the Figure \ref{figure:RC_BrG_spec}b. Lines are fit well by Gaussian profiles. In Table \ref{table:flux_compact}, we list the CRC sources, fluxes, and numbers of effective O7 stars ($N_{\rm O7}$) required for ionization of each source (region). In the table, we assume $N_{\rm lyc} = 10^{49}~\textrm{sec$^{-1}$}$ per O7 star \citep{2003ApJ...599.1333S}. The \Nlyc\ $\sim$ 1--3 $\times 10^{51}~\textrm{s$^{-1}$}$ for the compact thermal nebulae implies that there are over 100 effective O7 stars within each 8~pc region. Their high $T_b$ implies high $n_e$, consistent with their youth \citep{2006AJ....132.2383T}. The mass (M$_{cl}$) of the clusters needed to excite these \HII\ regions is 0.3--1 $\times 10^{5}~M_{\sun}$ (for an age $\lesssim$ 3 Myr, and a Kroupa Initial Mass Function (IMF) from 1--100 $M_{\sun}$).

Ages of the exciting clusters here are difficult to determine since they are not resolved from the background field stars. We estimate that clusters with masses $> 3 \times 10^{4}\,M_{\sun}$ are accountable for about 30\% -- 50\% star formation activity in a single burst if an initial cluster mass function with power index of -2, as found in Antennae systems \citep{1999ApJ...527L..81Z}, is assumed. These young clusters can excite compact \HII\ regions of a few pc in diameter. As time progresses, the compact \HII\ regions of these clusters expand into extended emission structures \citep[$\gtrsim$ 100 pc at age $>$ 5 Myr, as seen in the Antennae system;][]{1999AJ....118.1551W}. In NGC\,6946, we find about 15\% of the dereddened \BrG\ line flux in our field comes from these six cluster-excited \HII\ regions ($M_{\rm cluster} \gtrsim 3 \times 10^{4}\,M_{\sun}$). That is only 1/2 to 1/3 of the expected contribution from the compact \HII\ regions at zero age.  So we can limit the lifetime of \HII\ regions in the compact stage to about 1--1.5 Myr, 1/3--1/2 of the lifetime of a \BrG\ \HII\ region. This is similar to what is found in M82 \citep{2009AJ....137.4655T}.

\section{Discussion}\label{section:discussion}

\subsection{Star Cluster Formation in Circumnuclear Regions -- In situ or Migration}\label{section:SCF_nuclear_ring}

The stage of evolution in which a young cluster holds a significant population of massive ionizing stars and emits radio continuum and recombination lines is short, a few Myr. Therefore it is commonly assumed that any cluster observed via its ionized nebulae is close to its place of birth. But this assumption may not be a safe one near galactic centers, where rotational periods are short and velocities are high \citep[or, in regions with strong radial bar flow,][]{2009ApJ...703.1297E}. In galaxies with strong nuclear bars, there is an additional radial component of motion that scales with these speeded-up dynamical times. 

In NGC\,6946, we use the rotation curve of \cite{2006ApJ...649..181S} to estimate that the rotation period at $r=2\arcsec$ (55 pc) is 2.2 Myr, corresponding to an angular frequency $\Omega = \frac{v}{R} = 2800$~km/s/kpc. At $r=4\arcsec$, r = 107 pc, v=150~km/s, $\Omega \sim 1500$ km/s/kpc. However, the bar pattern is also rotating. \cite{2006ApJ...649..181S} estimate that the bar pattern speed is $\Omega_{p} \sim$ 500 -- 700 km/s/kpc. So the clusters are moving though the bar pattern at $\Omega - \Omega_{p} \simeq$ 2100 -- 2300 km/s/kpc at 2\arcsec\ (55 pc) or $\Omega - \Omega_{p} \simeq$ 800 -- 1000 km/s/kpc at r = 4\arcsec\ (110 pc). The stars will ``lap'' the pattern once every 3 Myr at $r = 55$ pc, and 7 Myr at $r = 4\arcsec$; they will rotate into the opposite arm emerging from the bar in half that time. Note also the differential rotation period between r = 2\arcsec\ and 4\arcsec, a distance of only 55~pc; tidal effects on large structures such as GMCs are important. 

Given these timescales, even young clusters can potentially be found far from their birth places,  and that strong tides of the galactic nucleus can be hostile to cluster formation and survival. Why do we find clusters in the nuclear region? Did they form in situ and are simply very, very young? Or have they migrated to their current locations in their short lifetimes? 

To examine this hypothesis and further study the formation environment of clusters near the nuclear region requires comparison of multiwavelength data with high astrometrical accuracy. In this work, the newly presented radio and \BrG\ data are combined with high-resolution CO maps and gas dynamics from literature \citep{2006ApJ...649..181S,2007A&A...462L..27S} to compare the dynamics of molecular gas with  cluster formation. 

NGC\,6946 hosts a nuclear molecular bar $\sim$ 300 pc in size along the north-south direction, observed in CO interferometric maps in 4\arcsec\ -- 8\arcsec\ resolution \citep{1990ApJ...355..436I,2004AJ....127.2069M}, which confirmed early single dish work \citep{1988PASJ...40..511S}. Sub-arcsecond CO(1-0) and CO(2-1) maps of the nuclear bar revealed structures \citep{2006ApJ...649..181S}, consistent with gas response to the potential of a smaller nuclear bar \citep{1995ApJ...452L..21R,1998AJ....116.1221E,2002MNRAS.329..502M,2006ApJ...649..181S}. 

For the dynamical picture of gas motion in the nuclear bar, we adopt the hydrodynamic model for the central kpc region of a singly barred galaxy by \cite[][, their Fig. 3]{2002MNRAS.329..502M}, which matches the gas morphology of NGC\,6946 remarkably well (Figure \ref{figure:sketch}) if the bar is scaled down to 300 pc extent. In this model molecular gas piles up on the leading edge of the nuclear bar, which features straight arms, offset from the nucleus (denoted the ``principal shock" --- PS in Figure \ref{figure:sketch}a). Gas densities are also high in the curved, trailing tail of the outer nuclear bar (marked the ``spiral shock" --- SS), and at the ``nuclear ring" (NR). In the clump of gas associated with the NR, circular gas motion is revealed by high resolution (0\farcs35) CO(2-1) observations \citep{2007A&A...462L..27S}. The clump, which is oriented close to perpendicular to the dust lanes, contains a total of $\sim 4 \times 10^{6}~M_{\sun}$ of molecular gas, or about 5\% of the dynamical mass within 2\arcsec\ region \citep[corrected to $X_{\rm CO}$ of NGC\,6946's nucleus;][]{2004AJ....127.2069M,2006ApJ...649..181S}. 

\input{fig10.tex}

The giant \HII\ regions B, C$^{\prime}$, D, and I are located on the NR (see Figure \ref{figure:Av_conti}), and are likely to be excited by young star clusters associated with the nuclear ring. It is reasonable to expect these clusters to have formed directly in the NR. Molecular gas columns are large and according to the \citet[][]{2007A&A...462L..27S} model, motions are primarily circular in the NR region. 

Sources E and H are less easily understood. These clusters are not located near the CO clumps with highest concentrations of molecular gas, but found between the outer CO peaks and the NR, where large peculiar motions exist and large velocity dispersions are expected. 

We discuss three possible birth scenarios for the clusters exciting sources H and E: ``\textit{inside-out scenario}'' -- they formed inside their current radius and have migrated outward; ``\textit{outside-in scenario}'', i.e., they formed in the dense gas just inside the outer CO peaks (where the nuclear PS and SS features meet), and are moving inward along the PS; or, they are currently very close to their birth place (``\textit{in-situ scenario}'', w.r.t. the stellar bar and gas). Figure \ref{figure:sketch}b--d shows the three scenarios. Both inside-out and outside-in scenarios allow the clusters associated with source E and H to have formed in more gas-rich environments than they are in at present. The migration is associated with noncircular motions resulting from the nuclear bar.

The inside-out model requires that either i) clusters form in the NR or ii) clusters form in the x$_1$-x$_2$ orbit intersection of the nuclear bar. The first possibility has the problem that the gas in the nuclear ring/spiral regions is likely to be on nearly circular orbits, having already lost much of its angular momentum, and the clusters born from this case would have little angular momentum to move very far outward. \citet[][]{2003ApJ...582..723R} show that outward migration is expected to be only a fraction of the ring radius. So this evolutionary scheme is unlikely to explain the locations of E and H. In the second possibility, the clusters could form in a short window of time with bulk flow of the molecular gas at the x$_1$-x$_2$ orbit intersection within the nuclear bar. Once they form, torques that operate on the gas may cease to operate on the stars and the clusters `coast' off in the direction of motion and drift into the `spray' region, finally reaching the leading edge of the nuclear bar where they are currently observed. Since the $x_{1}-x_{2}$ orbit intersection regions appear to be at a larger radius than the NR, much less radial motion is needed. This picture, as shown in the Figure \ref{figure:sketch}b, is a possible explanation, but both the roughly symmetric location of source E and H with respect to the \sersic\ center and the fact that they happen to align very closely with the present location of the PS on the opposite side seem somewhat fortuitous. It is also unlikely if the star formation efficiency is low and the clusters are dynamically influenced by their natal clouds. 

In the outside-in formation model, the clusters form at the ``convergence region" between the CO north and south clumps of the SS region and dust lanes of the PS regions, where models indicate that molecular gas has high gas density, smaller velocities and less dispersion \cite[][, marked ``C" in their Fig. 3]{2002MNRAS.329..502M}. This is a site that appears to be dynamically favorable to star formation. Once the clusters form, they are at the top of the PS, and the clouds subsequently stream down along the PS. In this scenario, the clusters partly inherit their radial motion from the inward flow of the natal gas. However, at standard star formation efficiencies of $\lesssim$10\%, if the clusters form embedded in the clouds, they will feel the influence of the cloud over the influence of the background stellar distribution out to a radius of $\sim$20--25~pc \citep[e.g.,][]{2008ApJ...675..281M}. In other words, the GMCs can ``assist" their young clusters to migrate radially down the PS toward the nucleus. To travel down the PS regions from the birthplaces to their current locations within a $\sim$3 Myr lifetime of an \HII\ region requires a velocity of $\sim$ 25 km/s, which is consistent with the observed rotation curve and bar models. As in the previous scenario, in the outside-in model there is no clear explanation for the geometric symmetry of the locations with respect to the dynamical pattern of nuclear gas and stellar bar, but the fact that they lie along the PS is explained.

The third scenario is that sources E and H formed in situ (i.e. forming at their current PS location). In this scenario, E and H must approximately co-move with their natal gas. So as to not separate from the molecular gas they must be extremely young, $< 1$~Myr. Argument against this scenario is that gas in the PS generally has a high velocity dispersion, so it is somewhat surprising that strong star formation would proceed there. Moreover, the radial velocities ($\pm$ 30 km/s) of these two giant \HII\ regions are lower than the CO velocities at the same radii. This might be an indication that the clusters which ionized the observed nebulae are detaching from their natal gas clouds or are on more circular orbits. This scenario seems to be the least likely, especially since two of the five \BrG-emitting regions are found in the PS. This implies that E and H cannot be too young unless we are catching the star formation at a very special time.

In summary, for the \HII\ regions E and H, we consider a) their formation at the x$_1$-x$_2$ orbit intersection and subsequent ``spray" to a larger radius (``inside-out" model) or b) formation in the ``convergence" region of high gas density and low gas velocity where the nuclear PS and SS features meet (``outside-in" model). The presence of favorable star formation conditions in the convergence region and the low SFE led us to favor the ``outside-in" model, in which the clusters are escorted from their birthplaces in the convergence region down the PS by their natal clouds. However, the roughly symmetric locations of source E and H with respect to the \sersic\ center are not automatically explained by either model.

\subsection{Star Formation Efficiency in Star Clusters}\label{section:SFE}

The overall star formation rate in the nuclear region of NGC\,6946 is high (0.2 -- 0.6 $M_{\sun}~\textrm{ yr$^{-1}$}$, from \BrG\ emission) but the formal SFE is not: \cite{2004AJ....127.2069M} derived SFE $< =$ 3\% over the inner 170 pc. This result, however, includes a large mass of molecular gas which is not involved in star formation at all. A more meaningful measure of the SFE would relate the masses of the young star clusters to the molecular clouds from which they formed. We use our high resolution spectroscopy to register clusters to their associated gas clouds and estimate the SFE in the natal clouds of clusters.

Figure \ref{figure:chan_Brg_CO} shows CO(2-1) overlay contours from \cite{2006ApJ...649..181S} on our \BrG\ map at the closest velocity channel. It is difficult to draw too close a comparison, since the intrinsic dispersions of the CO and \BrG\ lines are at least an order of magnitude different. The CO(2-1) channel maps have a channel width of 6~km/s, 5 times smaller than our \BrG\ channel width, so the CO(2-1) flux read from Figure \ref{figure:chan_Brg_CO} should be scaled up by 5 to match the \BrG\ velocity coverage. A major fraction of CO gas does not connect to any \HII\ emission (until the principal shock arms). 

Four clusters (sources B, C$^{\prime}$, D, and I) in the nuclear ring (NR) contain about a thousand O stars in total; the masses of the young clusters range from 6 -- 13 $\times 10^{4} ~M_{\sun}$, for a total $3 \times 10^{5}~M_{\sun}$, if a Kroupa IMF is assumed. The nuclear gas clump holds $\sim 4 \times 10^{6}~M_{\sun}$ of molecular gas. This suggests that the SFE in this region is about 10\% and 1/3 of the star formation happens in clusters of mass $M_{\rm cl} > 3 \times 10^{4}~M_{\sun}$. 

In the last two panels of Figure \ref{figure:chan_Brg_CO}, we highlight the CO channel maps at the central velocities of source E and H. These two sources are off the central \BrG\ complex, and we will derive their SFE with the assumption that they are still within their natal clouds. In the event the ``inside-out" scenario holds, this may not be the case. Source H contains about 60 O7 stars, with an estimated mass $\sim 0.2 \times 10^{5} M_{\sun}$. The corresponding peak in the CO(2-1) map is about 4.2 $Jy/beam ~ km ~ s^{-1}$, for a total $\sim 2.5 \times 10^{5} M_{\sun}$ $H_{2}$ mass in the 0\farcs35 beam ($\sim$ 10 pc in size). The SFE, defined as $M_{\rm cl}/(M_{\rm cl} + M_{\rm gas})$, is in excess of 7\%. Similar treatment of source E results in a SFE of $\sim$ 30\%. The latter value is at the higher end of the SFE in nearby embedded star clusters \citep{2003ARA&A..41...57L}. The high SFE value might be an argument against the picture that they are still in their birthplaces near their natal clouds (\S\ref{section:SCF_nuclear_ring}).

\subsection{Nuclear Cluster in Formation? Parallels with the Milky Way}\label{section:NSC}

The center of NGC\,6946 presents an interesting comparison to the Milky Way. Both galaxies are spirals, both have 0.5 -- 1.4 $\times 10^{9}~M_{\sun}$ in stellar mass within their central 300 pc \citep{2002A&A...384..112L,2004AJ....127.2069M} and similar gas content of 3 -- 5 $\times 10^{7} M_{\sun}$ in the same region \citep{1996ARA&A..34..645M,2004AJ....127.2069M}. The main difference is that the Milky Way has an earlier Hubble type and a larger central black hole \citep[$4 \times 10^{6}~M_{\sun}$ v.s. $< 10^{5}~M_{\sun}$; ][]{2007AAS...211.6906K} but not a nuclear bar. The Milky Way center harbors the Central Cluster, and the Arches and Quintuplet clusters are at 11\arcmin\ -- 12\arcmin\ (25 -- 30 pc in projection) from the center. The Galactic Central Cluster is a young cluster (4--7 Myr), but not very massive ($\sim 2 \times 10^{4}~M_{\sun}$). Because of its youth, it is luminous -- with a bolometric luminosity of $2 \times 10^{7}~L_{\sun}$ \citep{1995ApJ...447L..95K}, similar to luminosities of nuclear clusters in local spiral galaxies \citep{1998AJ....116...68C,2002AJ....123.1389B}. The Arches and Quintuplet clusters have similar mass, age, and luminosity to the Central Cluster \citep{2008arXiv0803.1619F}. How do the young nuclear clusters in NGC\,6946 compare?

\input{fig11.tex}

Unlike the most late-type spiral galaxies \citep{2002AJ....123.1389B}, NGC\,6946 does not have a distinct nuclear star cluster, or a ``central cusp'', in near-IR surface brightness. It is possible that a nuclear cluster in NGC\,6946 exists, but is obscured by the dust and gas in the central area \citep[][up to $A_{K} \sim 15$ mag]{2007A&A...462L..27S}. If the Galactic center clusters (Central, Arches, and Quintuplet) were in the dusty nuclear region of NGC\,6946, and more evolved (so as to be less detectable sources in recombination line emission), they would not be visible in near-IR. It is also possible, as \cite{2007A&A...462L..27S} suggest, that we may witness the formation of a nuclear cluster as the observed bar-driven molecular gas spiral structure concentrates the molecular gas on the 30 pc scale. As the $\sim 10^{6}~M_{\sun}$ gas mass accumulated in the center area converts into stars, the central cluster will become obvious. 

Our radio continuum and \BrG\ studies identify young \HII\ regions in the nuclear area of NGC\,6946. These young clusters are found within 30 pc of the dynamical center, and are more massive ($\sim 10^5\, M_{\sun}$) than the three ``massive'' star clusters in the Galactic Center of the Milky Way. However, they are not likely to migrate to the very center by dynamical friction as has been proposed for the Milky Way \citep{2003ApJ...597..312K}. In the simulation of \cite{2003ApJ...597..312K}, a cluster can survive to migrate to the central pc only if it forms close enough ($\lesssim$ 5 pc) to the center, and is massive enough ($> 10^{6}\,M_{\sun}$) to avoid the destruction by the dynamical friction of the ambient gravitational potential during inward migration. The simulation of cluster migration by \cite{2003ApJ...597..312K} fits the Milky Way, which hosts a supermassive black hole ($4 \times 10^{6} ~ M_{\sun}$) that NGC\,6946 lacks, but the conclusions apply for clusters orbiting well outside the black hole influence. 

The location of \HII\ regions E and H along the principal shocks (PS) of the bar suggests that these \HII\ regions are traveling inward following the inward flow of gas along the PS. At the relatively low star formation efficiencies seen, the GMCs could carry their young clusters very close into the nucleus before the clusters separate from the clouds. Eventually, the clusters will be separated from their natal clouds and will dissolve, thus building up the mass of the nuclear disk or the stellar bulge. This ``gas-assisted" migration by gas clouds orbiting in a bar potential, as may be the case for \BrG\ regions E and H, is an interesting possibility to explain how star clusters born well outside the nucleus can migrate inward and contribute to the formation of a compact nuclear cluster.

\section{Summary and Conclusions}\label{section:conclution}

Integral field spectroscopy of \BrG, \Htwo, and K-band continuum in the nucleus of the late-type spiral NGC\,6946 using the Keck OSIRIS instrument is presented. LGS-AO observing gives spatial resolutions of 0\farcs3 for the cubes, which corresponds to a resolution of $\sim$8 pc in the galaxy. By registering the \BrG\ line emission with radio continuum images of the \HII\ regions, we have achieved an absolute astrometry to 0\farcs1. With this high precision astrometry in multi-wavelength data we produce a high quality dereddend near-IR stellar continuum map, and register compact sources in images of radio continuum, IR continuum, \BrG\ line, and millimeter CO line emission.

The comparisons between the HST NICMOS images and OSIRIS images of stellar continuum and recombination lines reveal different extinction toward stars and \HII\ nebulae. The highest foreground screen K-band extinctions $A_K$ we observe are $\sim 2.7-3$ ($A_V \sim 25$) for both stellar continuum and recombination line, suggesting that we have reached the observational limit of optical depth and have not seen completely through the clouds. 

The dereddened stellar continuum distribution suggests a position angle (10$\degr$ -- 20$\degr$) slightly different from previous results on the nuclear bulge. The small scale IR-bright nuclear bulge ($r \sim$ 65 pc) is populated with stars $\lesssim$0.1 Gyr old, which contribute significantly in near-IR continuum emission. The fraction of the young population increases in the central 30 pc. The S\'ersic profile of the nuclear bulge is centered at 0\farcs5 away from the CO dynamical center. It could be the effect of a non-uniform extinction on stellar light.

The ratio between K-band \Htwo\ vibrational-rotational transition lines, $v =$1-0 S(1) and $v =$1-0 S(0), suggests that the collisional excitation in shocks dominates the excitation of H$_2$. CO and H$_2$ show poor spatial correlation beyond the central 50 pc region.

The \BrG\ line and radio continuum morphology corroborate that the enhanced star formation in the nuclear region of NGC\,6946 is due to bar-driven inflow of molecular gas from the disk. We find $N_{Lyc} \sim 5\pm 1 \times 10^{52}$ sec$^{-1}$ and a star formation rate of 0.2 -- 0.6 $M_{\sun}\,yr^{-1}$ for the central $R=200$\,pc region. One-third of the star formation activity is in the form of star clusters of mass $M_{\rm cl} > 2 \times 10^{4}~M_{\sun}$, confined to within $\sim$70~pc of the dynamical center of NGC\,6946. Four of the \BrG\ \HII\ regions are found within the ``nuclear ring" or ``nuclear spiral" region, at galactocentric radii $R< 50\, pc$. The high concentration of gas in the immediate nuclear region argues that they formed there. However, high densities of molecular gas exist at larger radii, out to $R\sim 150$~pc,  without obvious star formation. Two \HII\ regions are found within the principal shock regions of the bar: we argue that these star clusters are unlikely to have formed in situ. It is more likely that they formed further out, in the ``convergence region" of the nuclear bar, and have migrated to their current positions in the straight ``principal shock" regions by gas-assist from their natal GMCs. Through the process of GMC-assisted migration, star clusters can join molecular clouds in bar streaming that can build up a dense nuclear cluster.

\acknowledgments

The authors thank the anonymous referee for a very thorough and helpful review. CWT acknowledges Tuan Do for his help in OSIRIS observation. We are grateful for discussions on these results at a TIARA conference with Steve Maciejewski and Bruce Elmegreen. This work is based on data obtained at the W. M. Keck Observatory, which is operated as a scientific partnership among the California Institute of Technology, the University of California, and NASA, and was made possible by the generous financial support of the W. M. Keck Foundation. The authors also wish to recognize and acknowledge the very significant cultural role and reverence that the summit of Mauna Kea has always had within the indigenous Hawaiian community; we are most fortunate to have the opportunity to conduct observations from this mountain. The 18~cm radio data reported in this paper is based on archived data made with MERLIN, a National Facility operated by the University of Manchester at Jodrell Bank Observatory on behalf of STFC. 

{\it Facilities:} \facility{Keck (OSIRIS)}, \facility{VLA}, \facility{MERLIN}.\\

\input{references.tex}
\end{document}

%% file: table1.tex
\begin{deluxetable}{ll}
\tabletypesize{\scriptsize}
\tablewidth{0in}
\tablecaption{Basic Information of NGC\,6946\label{table:N6946}}
\tablehead{
\multicolumn{1}{c}{Parameter} &
\multicolumn{1}{c}{Value} 
}
\startdata
Revised Hubble Type\tablenotemark{a} 	& SAB(rs)cd				 	\\
$L_{\rm IR}$ 			& 	$\sim 2.2 \times 10^{9} ~L_{\sun}$ \\
Age of starburst\tablenotemark{b}				& 	7--20 Myr \\
Dynamical Center\tablenotemark{c}		& 20$^\textrm{h}$34$^\textrm{m}$52$^\textrm{s}$.305, +60\arcdeg09\arcmin14\farcs58	 \\
\textit{l}, \textit{b}		& 95\arcdeg.7, 11\arcdeg.7        	\\
\Vlsr					& 50 km\,s$^{-1}$			 \\
Adopted Distance\tablenotemark{d}		& 5.9 Mpc					 \\
Inclination Angle\tablenotemark{e}		& 40$\arcdeg$					\\
Position Angle\tablenotemark{e}			& 	242\arcdeg						\\
Nuclear Bulge\tablenotemark{f}		& 						  \\
\, \,  Center				& 20$^\textrm{h}$34$^\textrm{m}$52$^\textrm{s}$.253 ; +60\arcdeg09\arcmin14\farcs10		 \\
\, \,  \sersic\ index 			& 0.91						\\
\, \,  R$_e$				& 2\farcs18					\\
\, \,  Position Angle		& 153$\arcdeg$					\\
\, \,  $R_{maj}/R_{min}$		& 0.69
\enddata
\tablenotetext{a}{\cite{1991trcb.book.....D}}
\tablenotetext{b}{\cite{1996ApJ...467..227E}, based on model fitting of ground-based near-IR imaging data.}
\tablenotetext{c}{\cite{2006ApJ...649..181S}, from CO(2-1), typographical error corrected.}
\tablenotetext{d}{\cite{2000A&A...362..544K}, based on blue supergiants in satellites.}
\tablenotetext{e}{\cite{2007AJ....134.1827C}}
\tablenotetext{f}{This work.}
\end{deluxetable}

%% file: fig1.tex
\begin{figure*}  
\figurenum{1}
\epsscale{1}
\begin{center}
\plotone{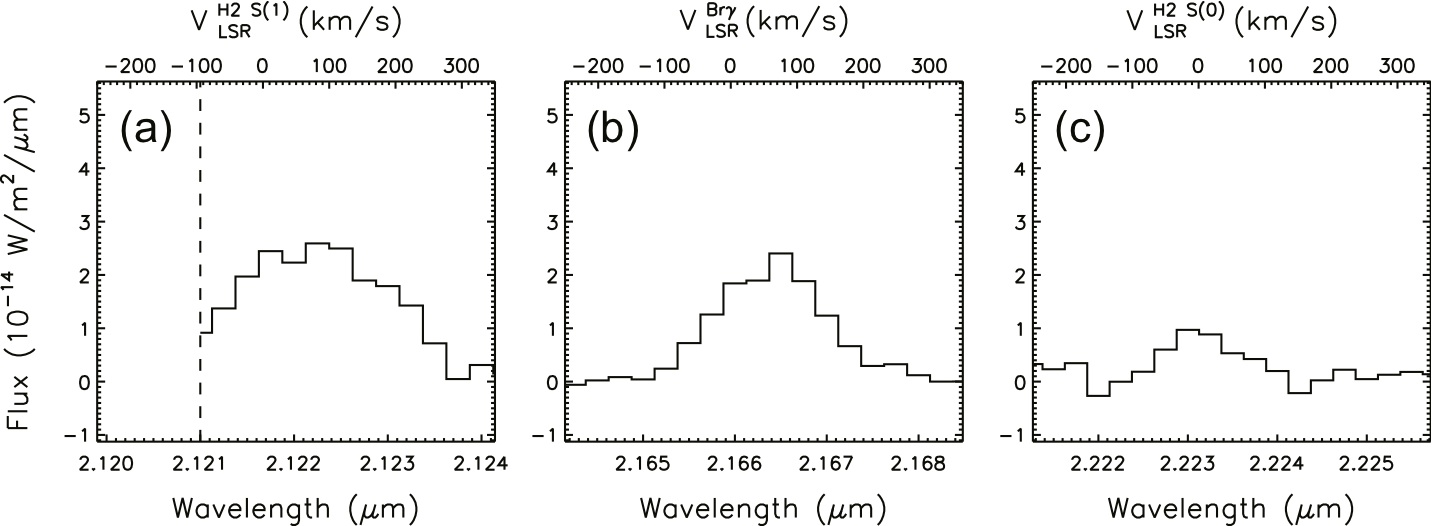}
\end{center}
\caption[Continuum-subtracted Integrated Spectra of Kn3 Band.]{The continuum-subtracted integrated spectra: (a) \Htwo\ 1-0 S(1), (b) \BrG, and (c) \Htwo\ 1-0 S(0) lines. The solid lines are the calibrated line profile over the entire emitting region (central 5\farcs9 $\times$ 2\farcs9, P.A. = 125$\degr$). The vertical dashed line marks the lower boundary of the wavelength coverage.}
\label{figure:spec}
\end{figure*}  

%% file: fig2.tex
\begin{figure*}  
\figurenum{2}
\epsscale{0.45}
\begin{center}
\plotone{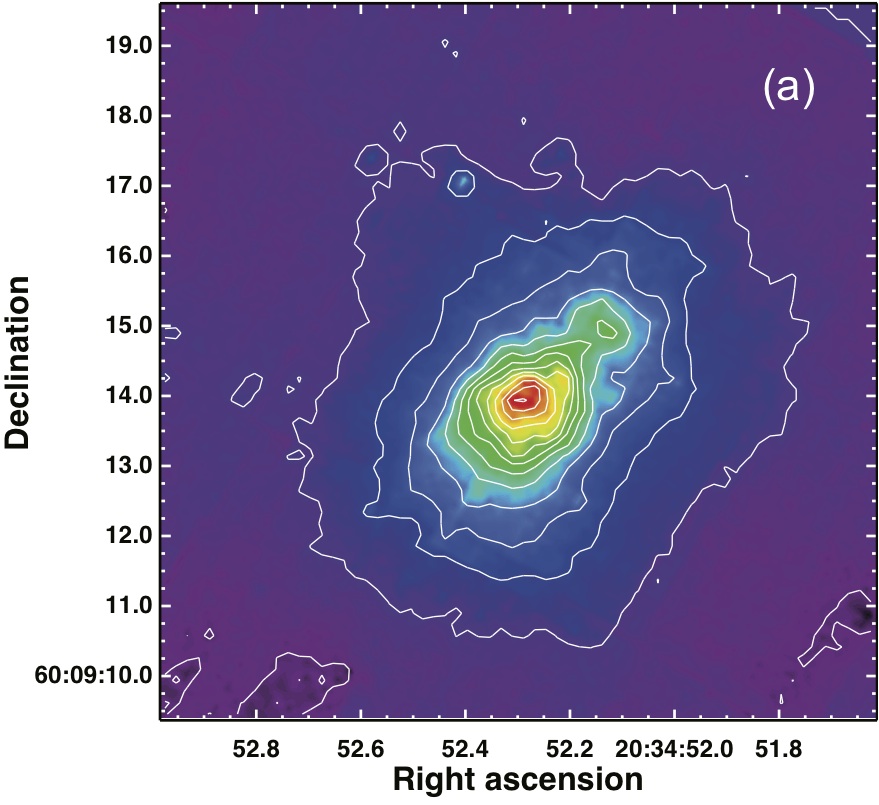}
\plotone{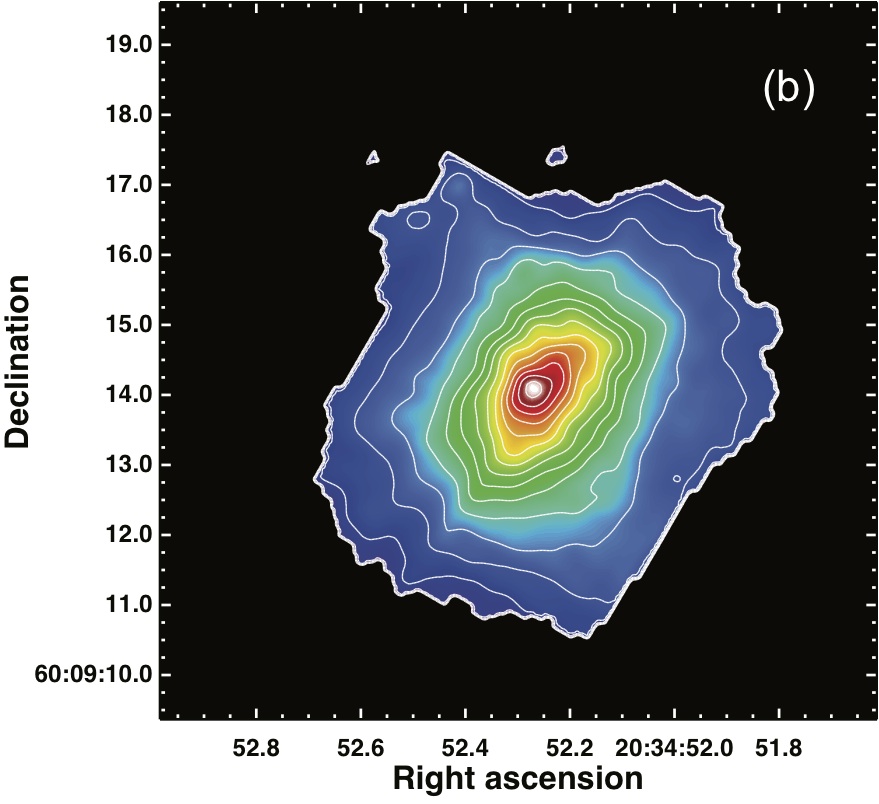}
\plotone{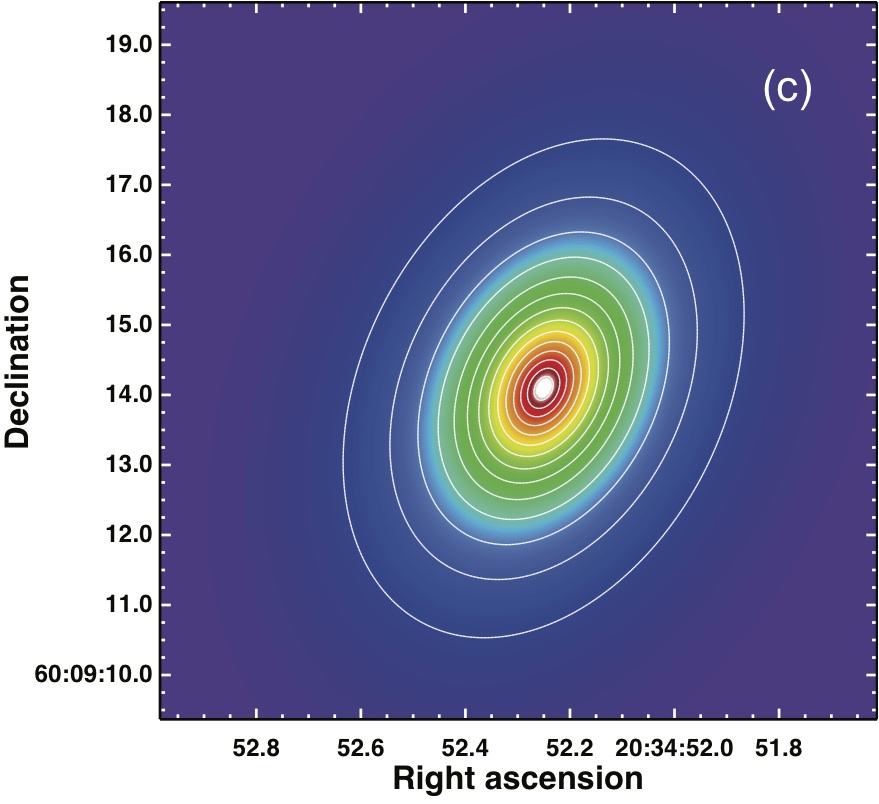}
\plotone{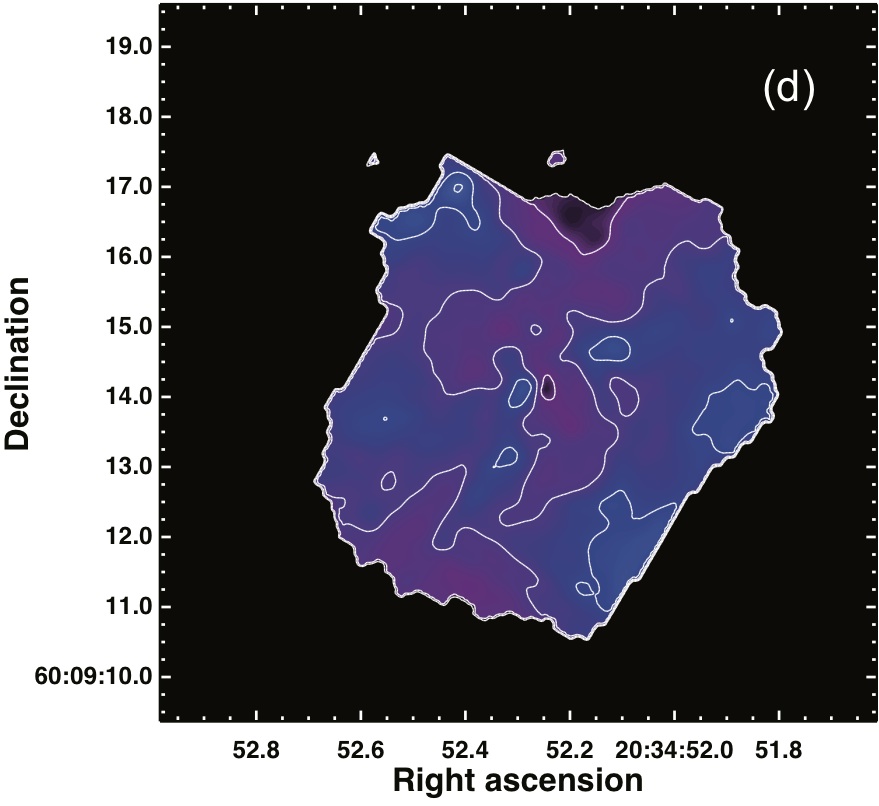}
\end{center}
\caption[Kn3-band Stellar Continuum Emission of NGC\,6946 nucleus]{(a) Kn3 band stellar continuum of NGC\,6946's nucleus. The peak is at $1.3\pm0.1 \times 10^{-14}~ W/m^{2}/\micron/arcsec^{2}$. The angular resolution is  0\farcs3. (b) De-reddened Kn3 stellar continuum map. (c) the S\'ersic model (defined in Section \ref{section:sersic_model}). (d) S\'ersic profile fitting residual. The color scales are the same for these four panels. The white contours are at  $1 \times 10^{-15}~ W/m^{2}/\micron/arcsec^{2}$.}
\label{figure:Kn3_obs}
\end{figure*}  

%% file: fig3.tex
\begin{figure}
\figurenum{3}
\epsscale{0.95}
\begin{center}
\plotone{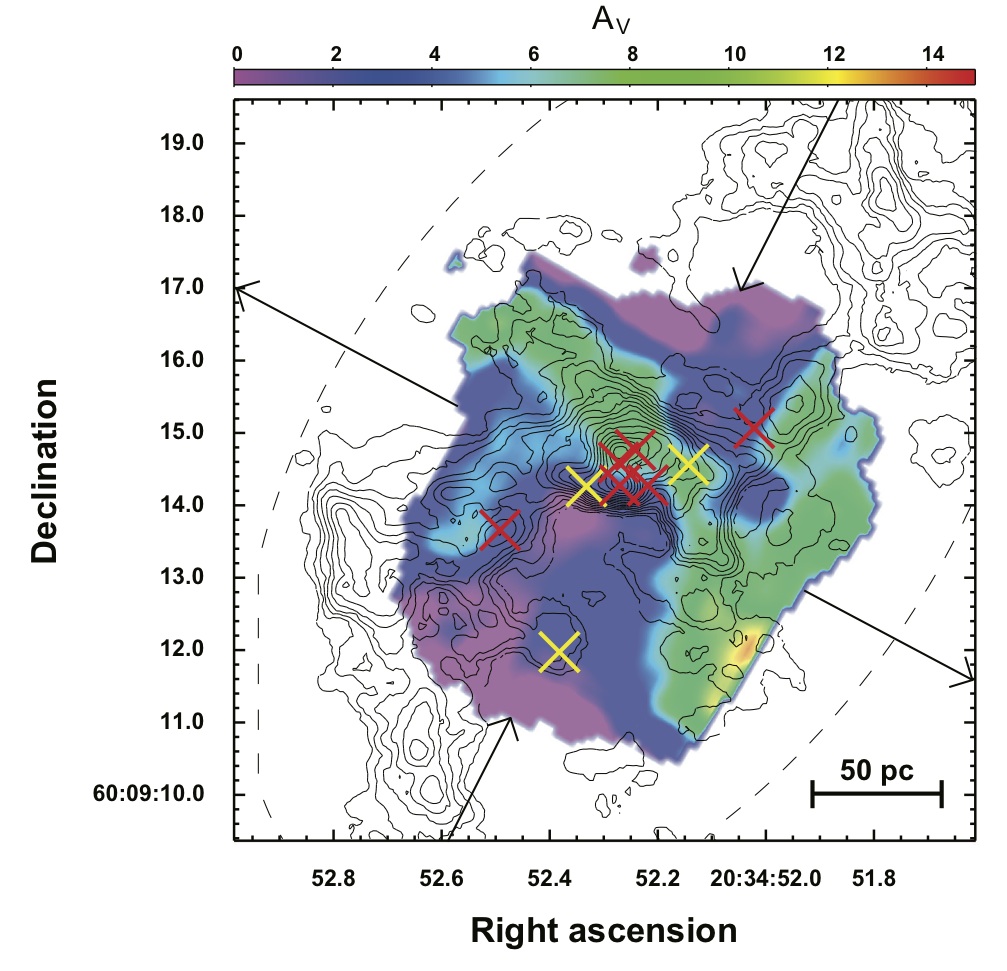}
\end{center}
\caption[Stellar Continuum Extinction Map.]{Stellar continuum extinction map. The colors indicate the $A_{{\rm V}}$ in magitudes. The contours are CO(2-1) line emission at 0.6~Jy/beam--km/s from \citet{2007A&A...462L..27S}. $A_{{\rm \tiny V}} = $ 10~mag equals to $I_{{\rm CO(2-1)}} = 1.3~{\rm Jy}~{\rm beam}^{-1}$, or $\sim$ two contour levels (see Section \ref{section:stellar_extinction} for detail). Thermal compact radio sources of \citet{2006AJ....132.2383T} are shown in red ``$\times$'' signs while the non-thermal sources are in yellow. The outward arrows point at the position angle of the large scale H\,\textsc{i} profile. The inward arrows indicate the orientation of the major axis of the dereddened stellar emission.}
\label{figure:Av_conti}
\end{figure}

%% file: fig4.tex
\begin{figure*}  
\figurenum{4}
\epsscale{0.45}
\begin{center}
\plotone{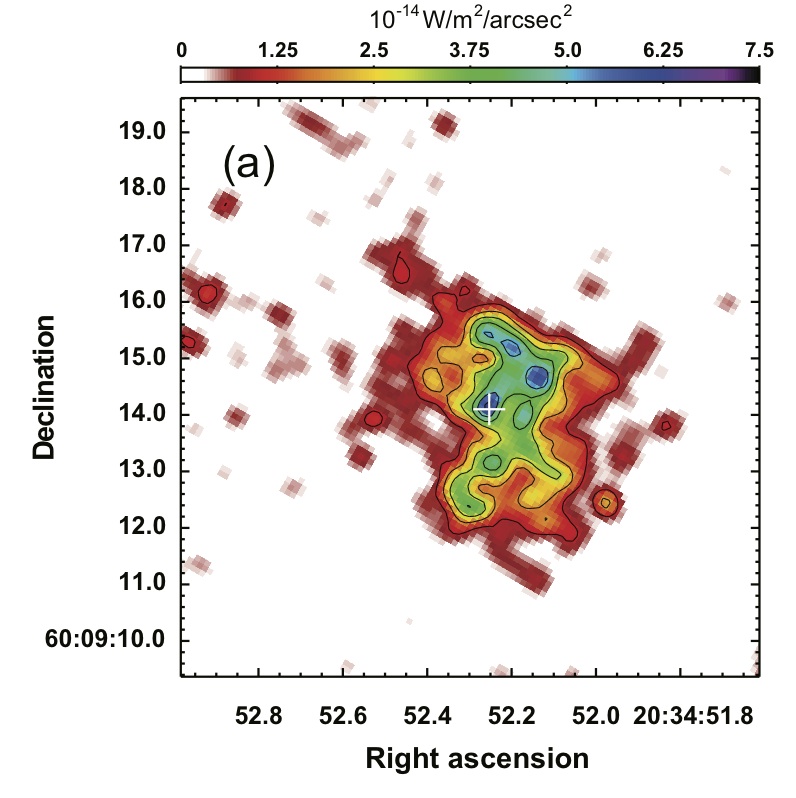}
\plotone{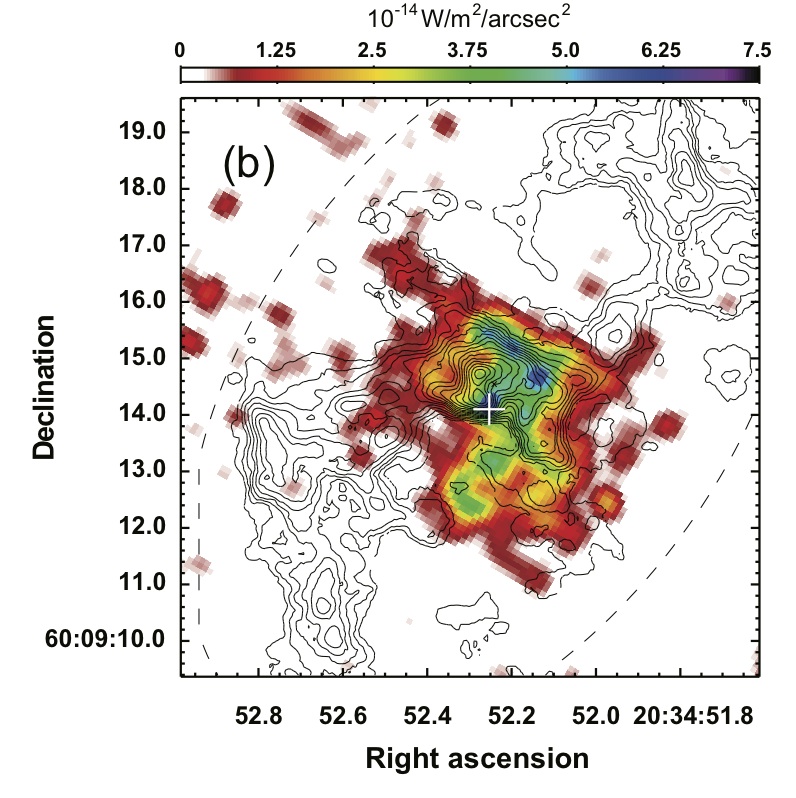}
\end{center}
\caption[Integrated $H_2$ 1-0 S(1) Line Map.]{(a) $H_2$ 1-0 S(1) line emission in color ($10^{-18}~W/m^2/arcsec^2$) and in contours (at $ 1 \times 10^{-18}~W/m^2/arcsec^2$ steps) (b) H$_2$ line map with CO(2-1) emission in contours. The CO(2-1) contours are at integers $\times$ 0.6 Jy/beam ~km/s from \citet{2006ApJ...649..181S}. The S\'ersic center is marked as white cross in both figures.}
\label{figure:H2_map}
\end{figure*}  

%% file: fig5.tex
\begin{figure}
\figurenum{5}
\epsscale{1.1}
\begin{center}
\plotone{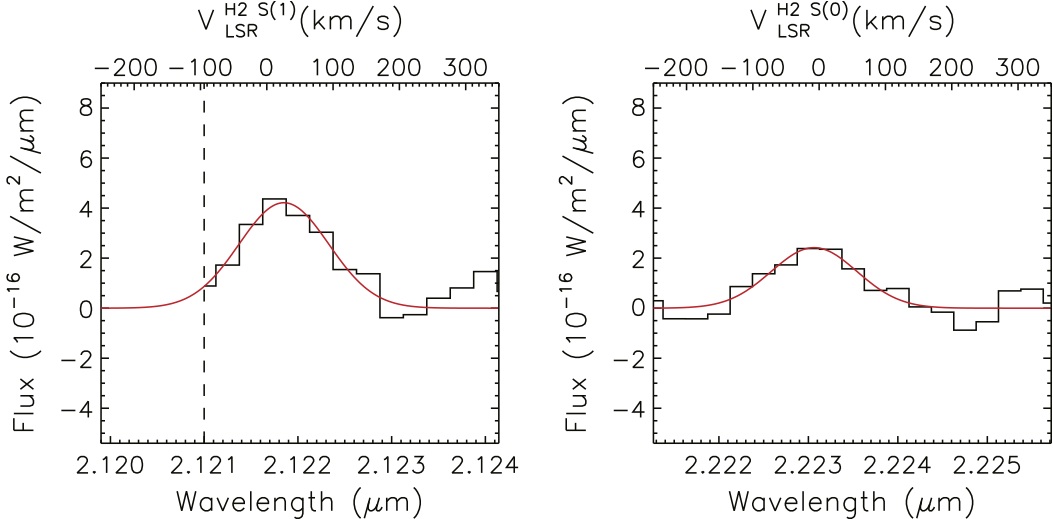}
\end{center}
\caption[Integrated continuum-subtracted H$_2$ 1-0 S(1) and 1-0 S(0) line Spectra of S\'ersic Center.]{Integrated H$_2$ 1-0 S(1) and 1-0 S(0) spectra over the 0\farcs3 region of S\'ersic center. The uncertainty of the spectra is at $0.38 \times 10^{-16}~W/m^2/\micron$. The solid red curves represent the best-fit Gaussian profile of the line.}
\label{figure:H2_spec_center}
\end{figure}

%% file: fig6.tex
\begin{figure}
\figurenum{6}
\epsscale{0.95}
\begin{center}
\plotone{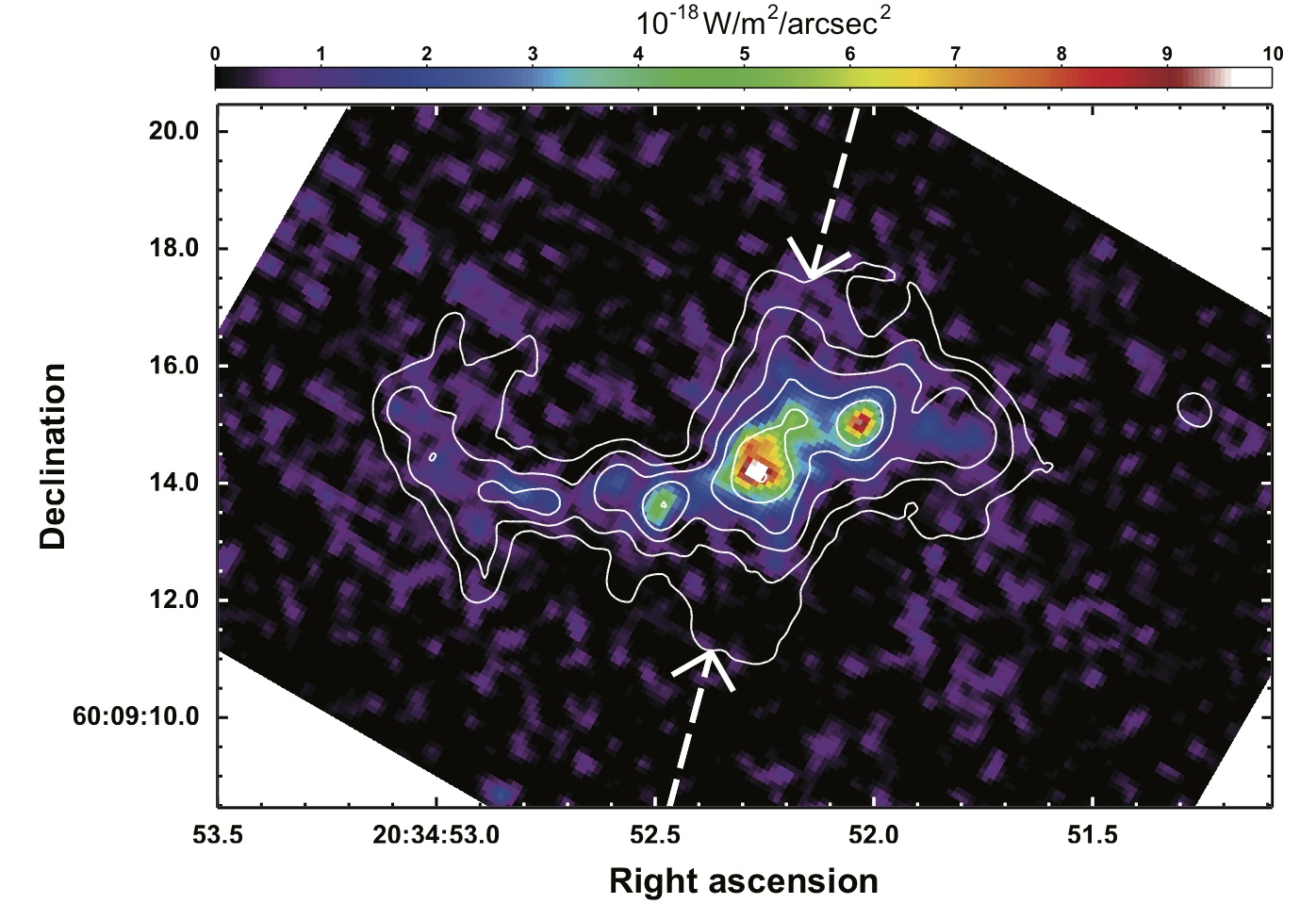}
\end{center}
\caption[Integrated $Br\gamma$ Line Map.]{Integrated $Br\gamma$ line map. The values of the color bar are in units of $10^{-18}~ W/m^{2}/arcsec^{2}$. The image is convolved to $0.3\arcsec$ PSF. The uncertainty of the line map is $0.4 \times 10^{-18}~ W/m^{2}/arcsec^{2}$. The contours outline NICMOS NIC3 $Pa\alpha$ emission at $2^{n} \times 10^{-18}~ W/m^{2}/arcsec^{2}$ with linear contours. The arrows with dashed lines point to an elongated \PaA\ emission component which is not clearly detected in the integrated \BrG\ image.}\label{figure:BrG_PaA}
\end{figure}

%% file: fig7.tex
\begin{figure*}  
\figurenum{7}
\epsscale{1}
\begin{center}
\plotone{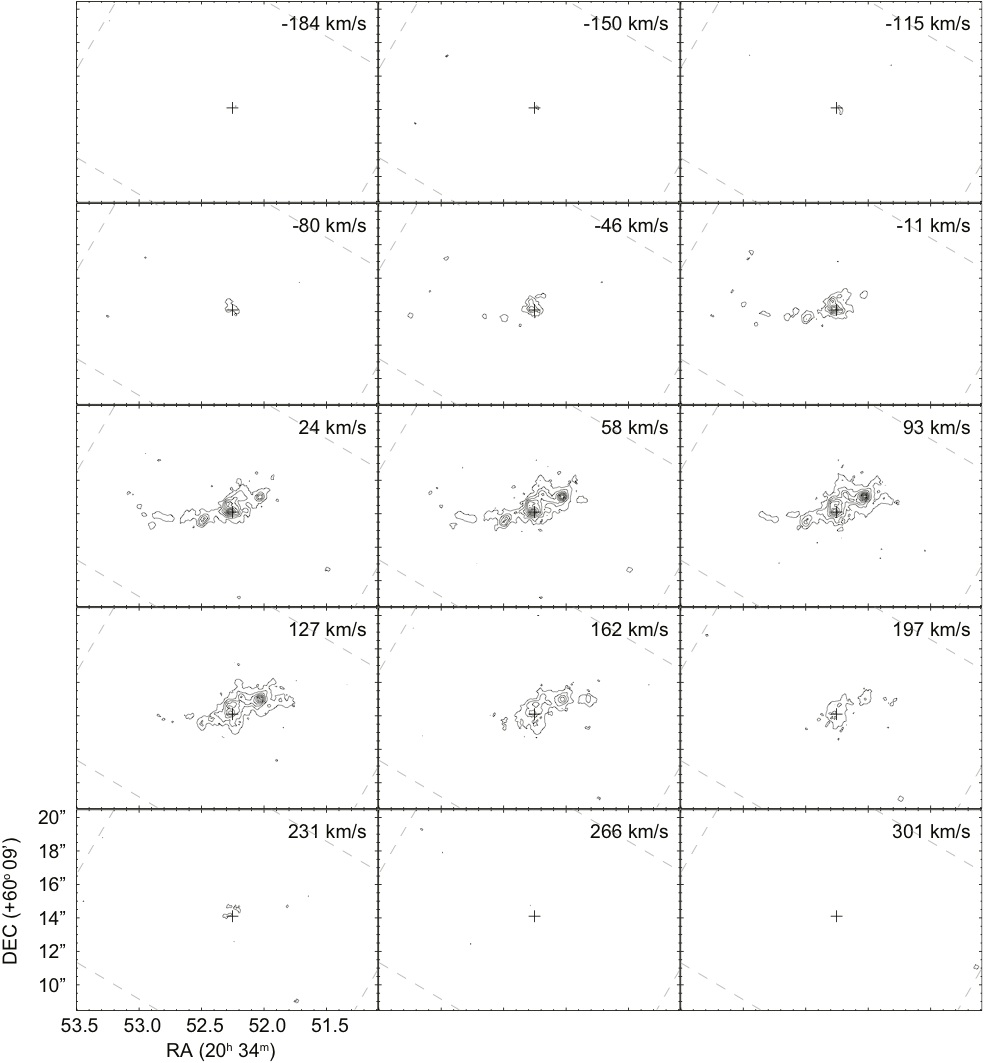}
\end{center}
\caption[$Br\gamma$ Line Channel Maps.]{$Br\gamma$ line channel maps after 2-dimension 3-point Hanning smoothing in the spatial domain. Each channel of the OSIRIS cube covers $0.25~nm$ ($\sim 35 ~ km/s$). The central $V_{\rm LSR}$ of each channel is showing in the \textit{top right corner} of each panel. The systemic velocity is $\sim +50$ km/s with respect to $V_{\rm LSR}$. The cross indicates the center of the best-fit S\'ersic model from \textsc{ galfit} (see Section \ref{section:sersic_model}). Contour levels are at positive integers $\times 10^{-15}~W/m^{2}/\micron/arcsec^{2} \sim 3 \sigma$.}
\label{figure:Chan_map}
\end{figure*}  

%% file: table2.tex
\begin{deluxetable*}{lcccrclclclc}  
\tabletypesize{\scriptsize}
\tablewidth{0in}
\tablecaption{Compact Source Observed Quantities\label{table:flux_compact}}
\tablehead{
\multicolumn{1}{l}{Source} &
\multicolumn{1}{c}{$\alpha$(J2000)} &
\multicolumn{1}{c}{$\delta$(J2000)} &
\multicolumn{1}{c}{$S_{6 cm}^{Peak}$\tablenotemark{;a}} &
\multicolumn{1}{c}{$S_{2 cm}^{Peak}$\tablenotemark{;a}} &
\multicolumn{1}{c}{$S_{Br\gamma}^{Obs}$} &
\multicolumn{1}{c}{$A_{Br\gamma}$} &
\multicolumn{1}{c}{$S_{Br\gamma}^{De-reddened}$} &
\multicolumn{1}{c}{$N_{O7}(Br\gamma)$\tablenotemark{b}} &
\multicolumn{1}{c}{$N_{O7}(RC)$\tablenotemark{b}} &
\\
\colhead{} &
\colhead{20$^{h}$34$^{m}$+ ($^{s}$)} &
\colhead{+60$\arcdeg$09$\arcmin$ + ($\arcsec$)} &
\colhead{(mJy beam$^{-1})$} &
\colhead{(mJy beam$^{-1})$} &
\colhead{$10^{-18}~W/m^2$} &
\colhead{mag} &
\colhead{$10^{-18}~W/m^2$}
}
\startdata
A\tablenotemark{c} & 52.34 & 14.2 & 1.5 & 1.1         &  0.50  & 2.78  & 6.49    & 86 & 330\tablenotemark{d}  \\
B & 52.28 & 14.2 & 1.4 & 1.5         &  1.09  & 2.78  & 14.15  & 186 & 450  \\
C$^{\prime}$\tablenotemark{e} & 52.29 & 14.6 & 1.1 & 0.7         &  0.81  & 2.34  & 7.00    & 92 & 210\tablenotemark{d}  \\
D & 52.25 & 14.7 & 1.0 & 1.0         &  0.64  & 2.55  & 6.70    & 88 & 300  \\
E & 52.03 & 15.0 & 0.3 & 0.5         &  0.83  & 2.32  & 7.01    & 93 & 150  \\
F\tablenotemark{c} & 52.39 & 11.9 & 0.2 & $<$ 0.4  & ---      & ---      & ---       & ---     & 60\tablenotemark{d}  \\
G\tablenotemark{c} & 52.15 & 14.5 & 0.2 & $<$ 0.4 &  0.28  & 2.22  & 2.17  &  29 &  60\tablenotemark{d} \\
H & 52.50 & 13.6 & 0.1 & $<$ 0.4 &  0.49  & 2.48  & 4.81  & 63 & 30\tablenotemark{d}  \\
I & 52.23 & 14.2 & 0.4  & 1.3          &  0.71  & 2.27  & 5.75  & 77 & 450  
\enddata
\tablenotetext{a}{$S_{6 cm}^{Peak}$ and $S_{2 cm}^{Peak}$: peak flux density (intensity) of 6~cm and 2~cm continuum, respectively \citep{2006AJ....132.2383T}.}
\tablenotetext{b}{Following Eq. (\ref{eq:Nlyc_from_BrG}) in Section \ref{section:Av_line}. Assume $N_\textrm{Lyc} = 10^{49}~\textrm{sec$^{-1}$}$ per effective O7 star \citep{2003ApJ...599.1333S}.}
\tablenotetext{c}{Suspected non-thermal source. Fluxes are measured in the 0\farcs3 region. No Br\,{$\gamma$} emission detected on source F. }
\tablenotetext{d}{From 6~cm flux density, assuming flat free-free spectrum ($\alpha = -0.1$). If source is optically thick at 6 cm, the number is the lower limit.}
\tablenotetext{e}{The corrected source C. See Section \ref{section:compact_objects} for detail.}
\end{deluxetable*}  

%% file: fig8.tex
\begin{figure*}  
\figurenum{8}
\epsscale{1.1}
\begin{center}
\plottwo{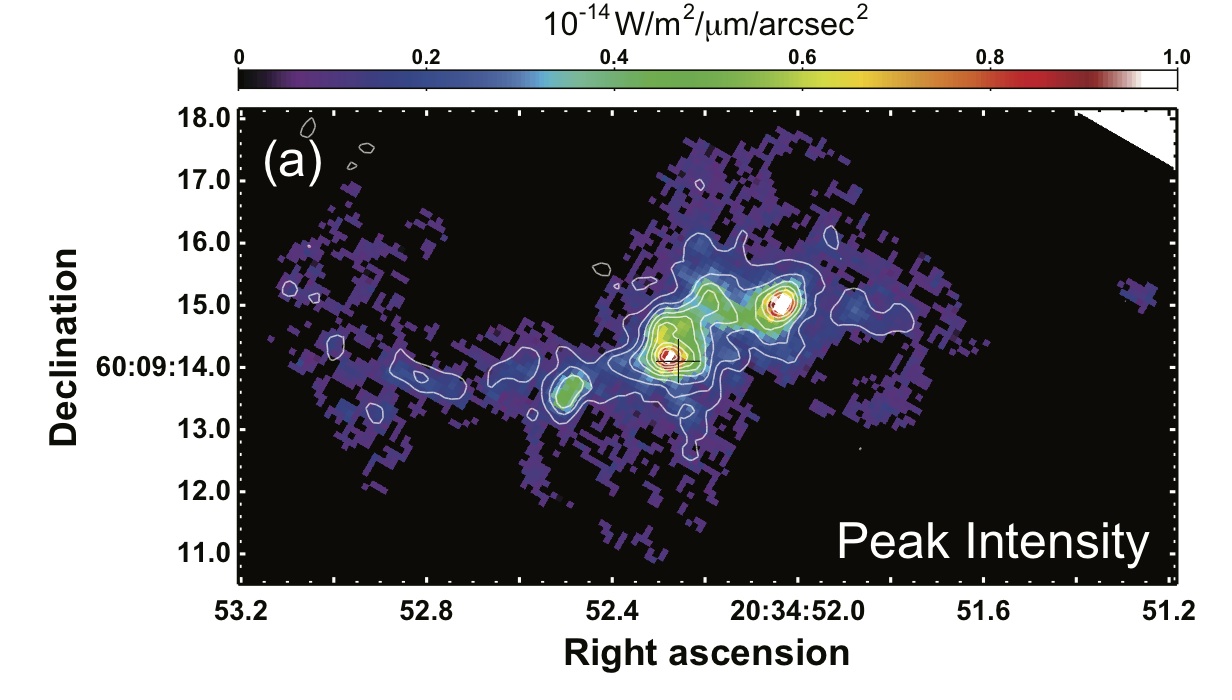}{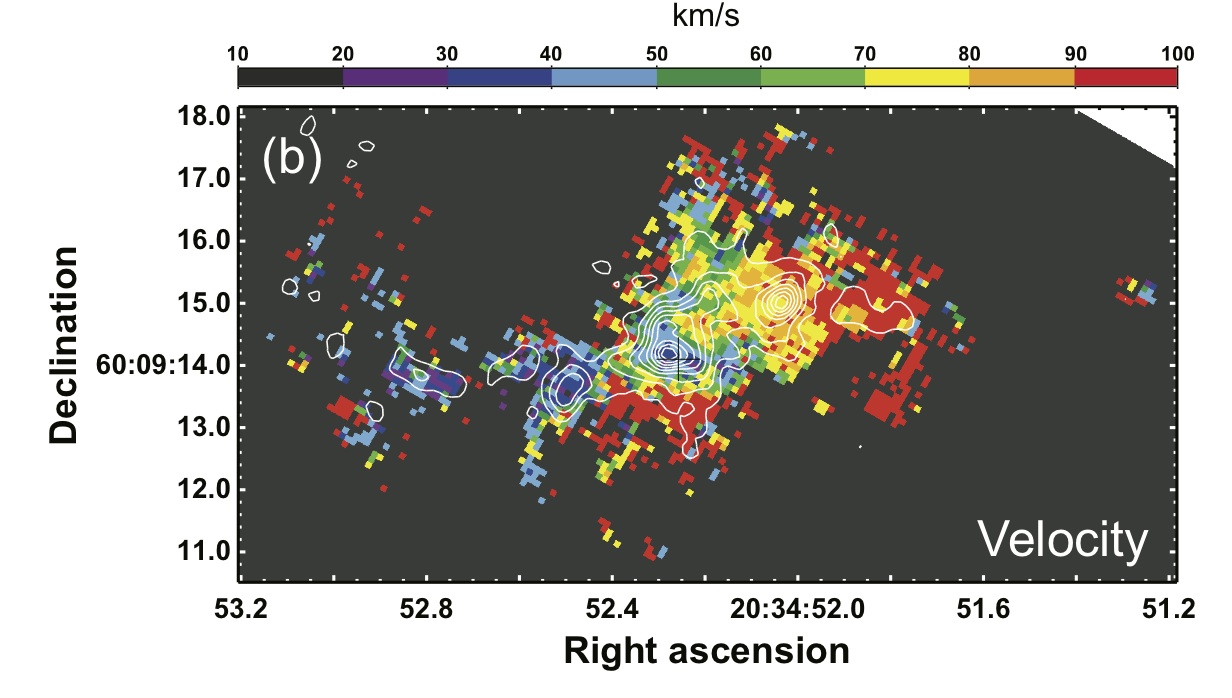}
\plottwo{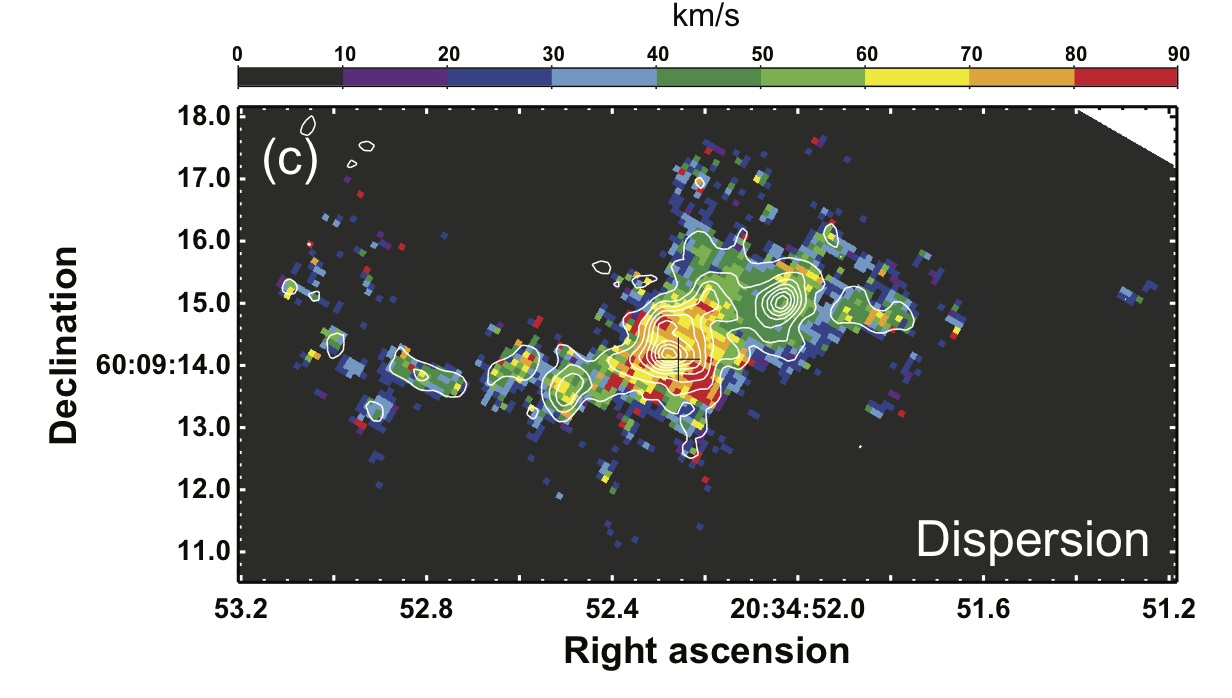}{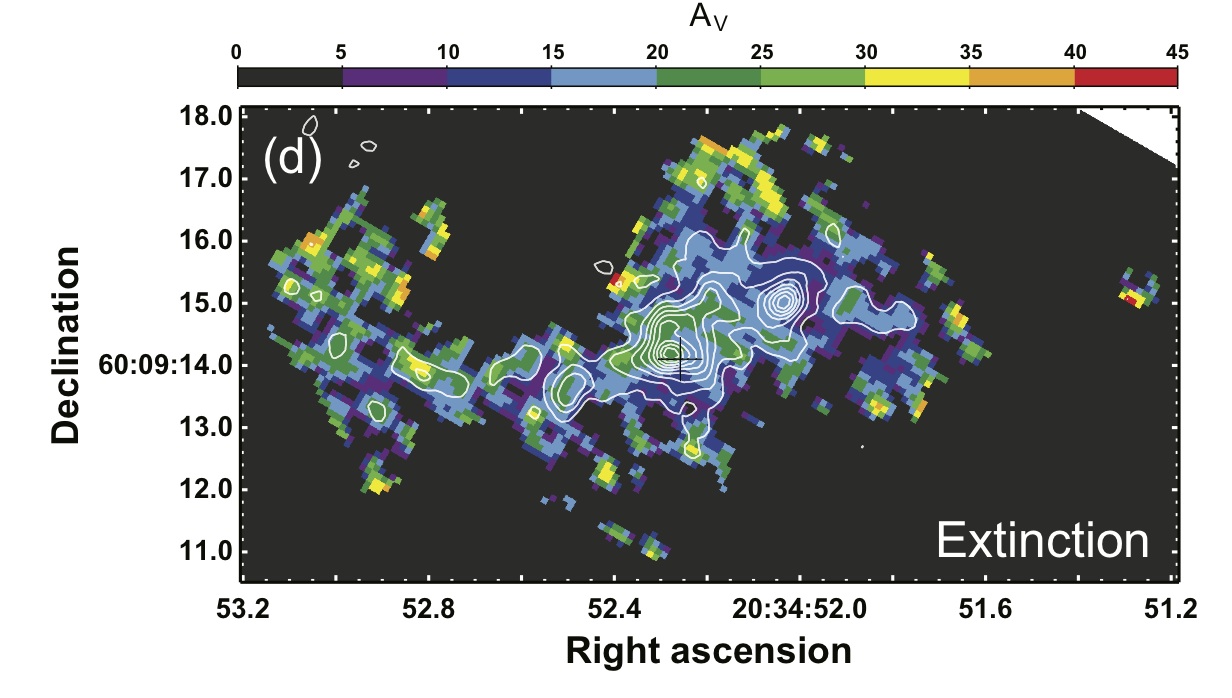}
\end{center}
\caption[$Br\gamma$ Line Fitting Maps and Extinction Map.]{$Br\gamma$ line (a) peak intensity map, (b) velocity field, (c) dispersion map, and (d) extinction ($A_{V}$) map. First three maps are results of gaussian line fitting at $0.1\arcsec \times 0.1\arcsec$ pixels. Peak flux map scale ranges 0--1 $\times 10^{-14}~ W/m^{2}/\micron/arcsec^{2}$. The velocity given in (b) is relative to LSR. The velocity is shown in $km/s$ in (b). S\'ersic fitting center is shown as black cross in each panel. Contours represent the integrated $Br\gamma$ line emission at levels the same as shown in Figure \ref{figure:Kn3_obs}. All maps are convolved to $0.3\arcsec \times 0.3\arcsec$ PSF. The dispersion is shown in $km/s$ in (c). The $A_{V}$ in magnitude is shown in extinction map (d), produced from $Br\gamma/Pa\alpha$.}
\label{figure:BrG_fitting}
\end{figure*}  

%% file: fig9.tex
\begin{figure}
\figurenum{9}
\epsscale{1.2}
\begin{center}
\plotone{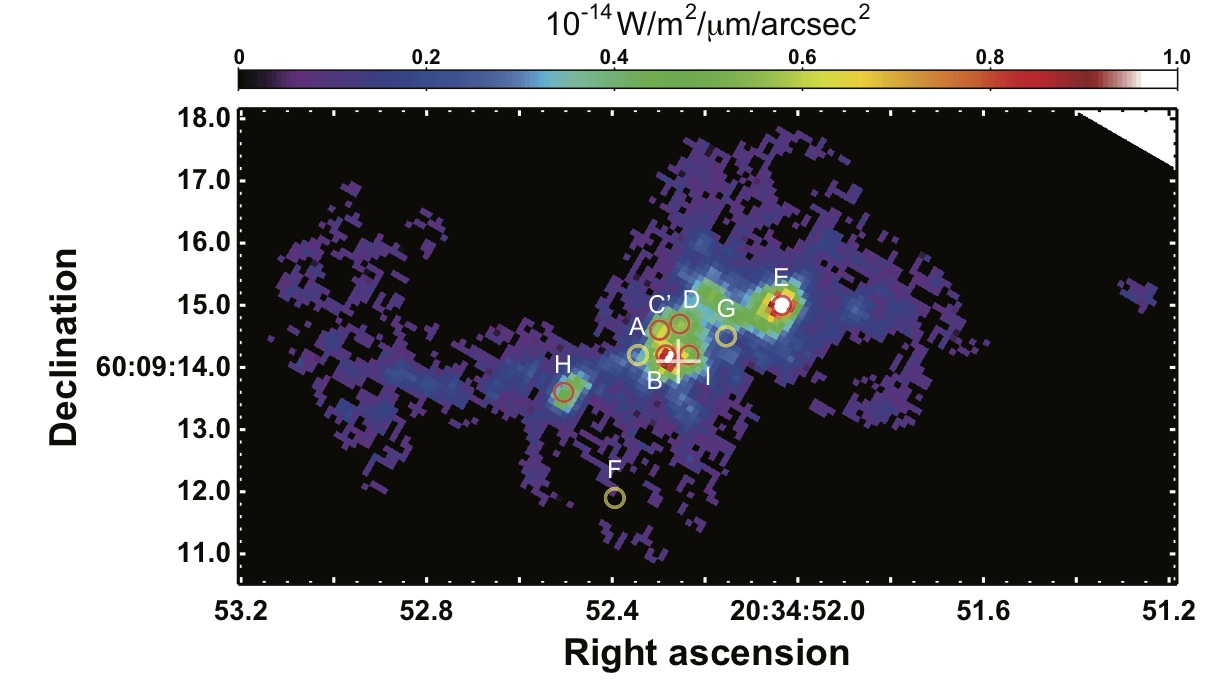}
\end{center}
\caption{(a) Radio continuum sources in the $Br\gamma$ peak intensity map. Compact radio continuum sources are sorted alphabetically by their 6~cm flux density in $\sim 0.4\arcsec$ beam \citep{2006AJ....132.2383T}. Circles, which are $0.3\arcsec$ in diameter, are thermal radio H\,\textsc{ii} regions in red and non-thermal supernova remnants in yellow. Cross is the \sersic\ center from \textsc{galfit} modeling.}
\label{figure:RC_BrG_spec}
\end{figure}

\begin{figure}
\figurenum{9}
\epsscale{1.2}
\begin{center}
\plotone{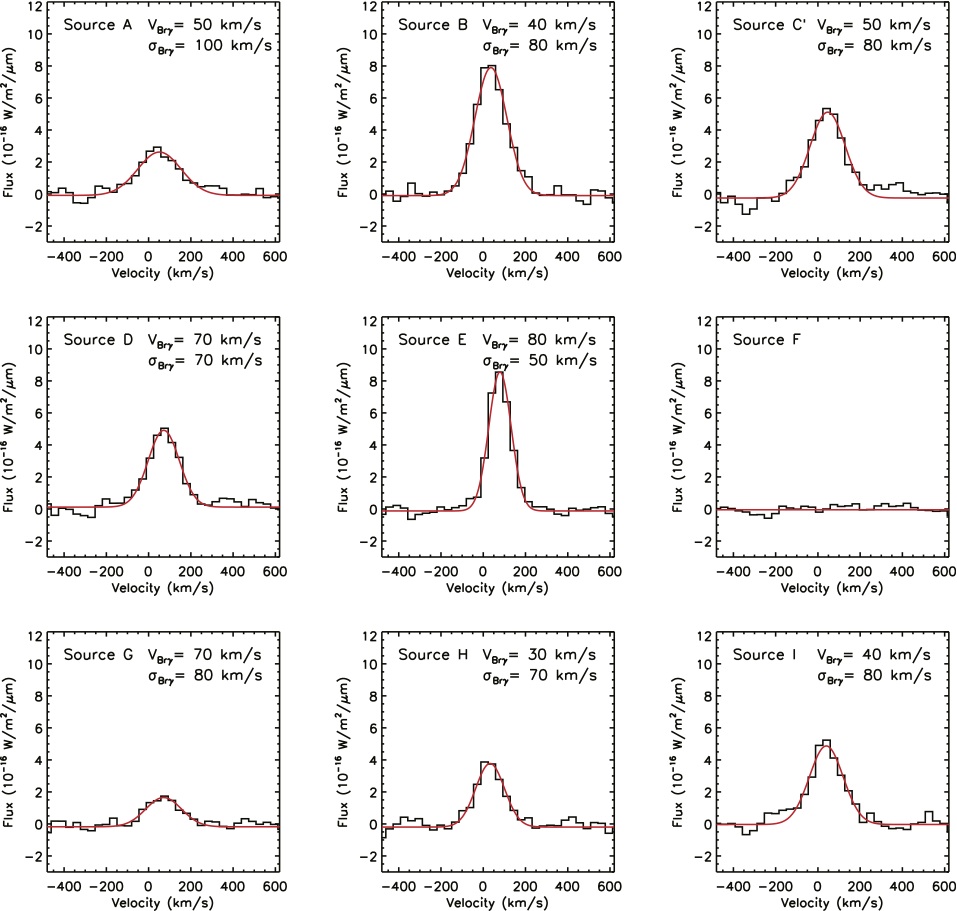}
\end{center}
\caption[Continued.]{Continued. The continuum-subtracted $Br\gamma$ line profiles of all nine radio continuum sources are displayed in (b). The spectra are integrated over the $0.3\arcsec$ diameter regions. The uncertainty on the individual spectrum is $0.33 \times 10^{-16}~W/m^2/\micron$. The velocity is relative to LSR. The typical uncertainty in central velocities and velocity dispersion is $<$ 10 km/s. The depletion feature around 250 km/s is due to atmosphere OH absorption.}
\end{figure}

%% file: fig10.tex
\begin{figure*}  
\figurenum{10}
\epsscale{1}
\begin{center}
\plotone{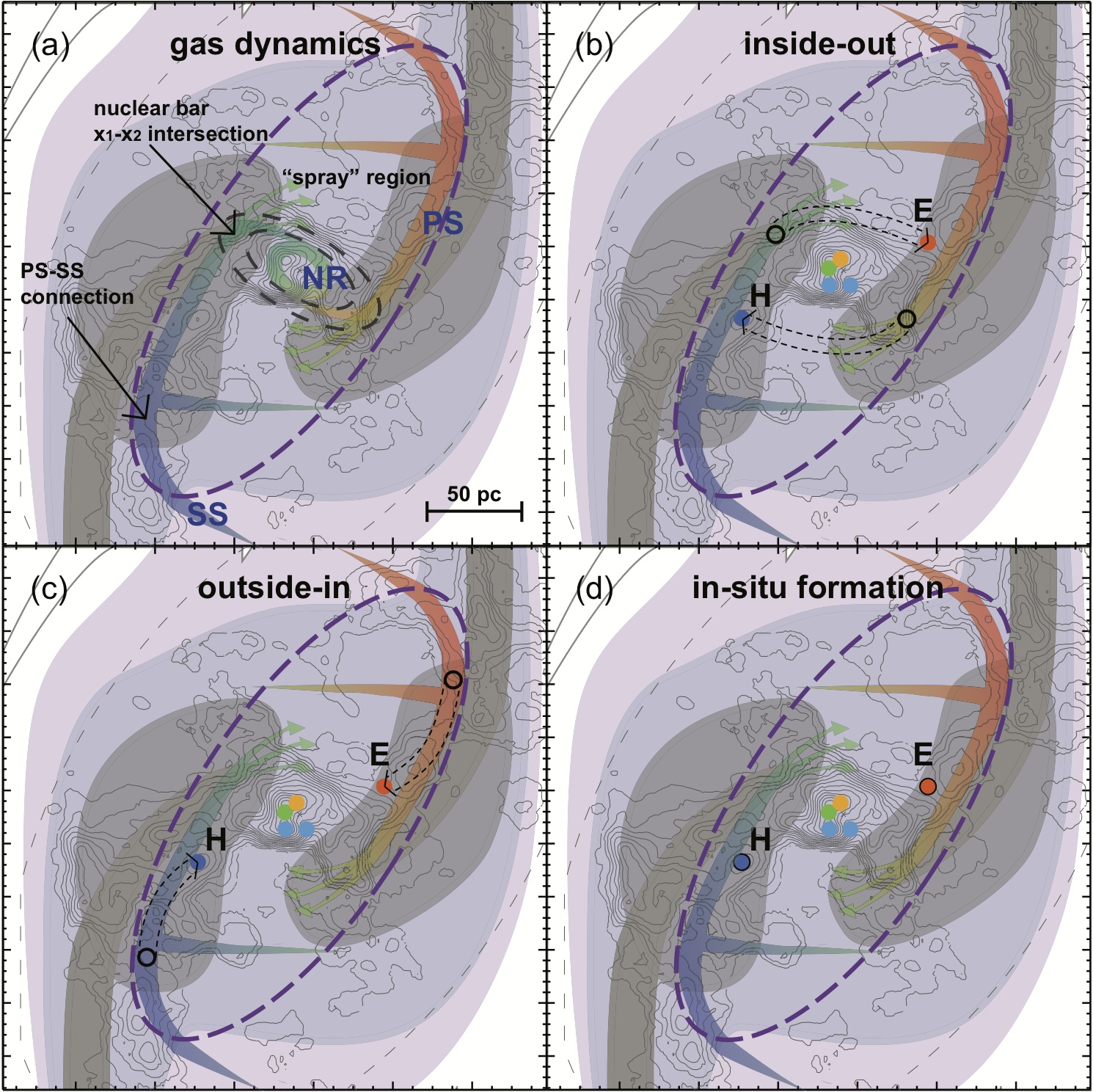}
\end{center}
\caption{Schematic showing (a) a plausible dynamical configuration of molecular gas under the influence of the nuclear bar, and the 3 scenarios of birthplaces of sources E and H under the influence of molecular gas motion: (b) \textit{inside-out}, (c) \textit{outside-in}, and (d) \textit{in-situ}. The filled circles in (b)--(d) are the observed \HII\ regions of young clusters. The black circles are possible birth places of clusters associated with CRC source E and H. The arrows represent gas flows and cluster motion with the nuclear bar. The colors of gas flows and \HII\ regions indicate the projected velocity on the line of sight. Contours show the CO(1-0) integrated intensity from \cite{2006ApJ...649..181S}. The background grey and purple regions adopted from \cite{2004AJ....127.2069M} illustrate the gas density associated with the larger bar. The schematic arrows illustrate the gas motion in the high surface density area, based on the doubly barred galaxy gas dynamics model by \cite{2002MNRAS.329..502M}. From the outside the gas ``tail'' trailing from the end of the nuclear bar potential indicates the region of the Spiral Shock (SS). The two gas arms on the leading edge of the nuclear bar and following the bar inward are the Primary Shock (PS) areas. At the very center the Nuclear Ring is marked as (NR). See Section \ref{section:SCF_nuclear_ring} for discussion.}
\label{figure:sketch}
\end{figure*}  

%% file: fig11.tex
\begin{figure*}  
\figurenum{11}
\epsscale{1.0}
\begin{center}
\plotone{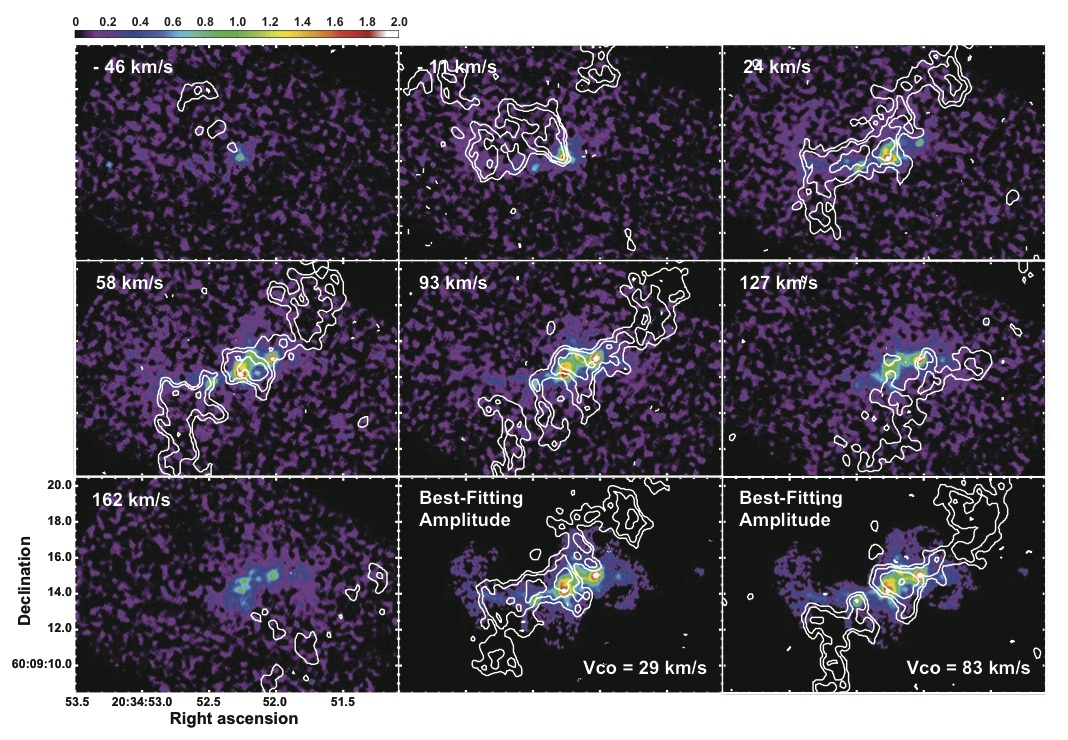}
\end{center}
\caption[Channel Maps of Br\,{$\gamma$} Line and CO(2-1) Line Emission.]{Channel maps of Br\,{$\gamma$} line from $V_{\rm LSR}$ = -50 km/s to $V_{\rm LSR}$ = 160 km/s. The contours are CO(2-1) line emission intensity from \citet{2006ApJ...649..181S} at $2^{N} \times 36~{\rm mJy/beam}$ ($N$ is natural numbers) in the closest velocity channel. The velocity difference is less than 3 km/s. The Br\,{$\gamma$} channel width is 35~km/s while the CO(2-1) channel width is 6~km/s. The last two color images are the same as Figure \ref{figure:BrG_fitting}a, and the CO(2-1) contours are at channels near the central velocities of source E ($V_{\rm LSR}$ = 80 km/s) and H ($V_{\rm LSR}$ = 30 km/s).}
\label{figure:chan_Brg_CO}
\end{figure*}  